\documentclass[10pt,preprint]{aastex}
\usepackage{epsfig}  

\newcommand{\etal}{et~al.}
\newcommand{\HeI}{\hbox{{\rm He}\kern 0.1em{\sc i}}}
\newcommand{\HeII}{\hbox{{\rm He}\kern 0.1em{\sc ii}}}
\newcommand{\HI}{\hbox{{\rm H}\kern 0.1em{\sc i}}}
\newcommand{\HII}{\hbox{{\rm H}\kern 0.1em{\sc ii}}}
\newcommand{\CIV}{\hbox{{\rm C}\kern 0.1em{\sc iv}}}
\newcommand{\NII}{\hbox{{\rm N}\kern 0.1em{\sc ii}}}
\newcommand{\SiIV}{\hbox{{\rm Si}\kern 0.1em{\sc iv}}}
\newcommand{\OI}{\hbox{{\rm O}\kern 0.1em{\sc i}}}
\newcommand{\OII}{\hbox{{\rm O}\kern 0.1em{\sc ii}}}
\newcommand{\OIII}{\hbox{{\rm O}\kern 0.1em{\sc iii}}}
\newcommand{\FeII}{\hbox{{\rm Fe}\kern 0.1em{\sc ii}}}
\newcommand{\NeII}{\hbox{{\rm Ne}\kern 0.1em{\sc ii}}}
\newcommand{\NeIII}{\hbox{{\rm Ne}\kern 0.1em{\sc iii}}}
\newcommand{\ArII}{\hbox{{\rm Ar}\kern 0.1em{\sc ii}}}
\newcommand{\ArIII}{\hbox{{\rm Ar}\kern 0.1em{\sc iii}}}
\newcommand{\SII}{\hbox{{\rm S}\kern 0.1em{\sc ii}}}
\newcommand{\SIII}{\hbox{{\rm S}\kern 0.1em{\sc iii}}}
\newcommand{\SIV}{\hbox{{\rm S}\kern 0.1em{\sc iv}}}
\newcommand{\Halpha}{\hbox{{\rm H}\kern 0.1em$\alpha$}}
\newcommand{\Hbeta}{\hbox{{\rm H}\kern 0.1em$\beta$}}
\newcommand{\Heopta}{\hbox{{\rm He}\kern 0.1em{\sc i}}~$6678$}
\newcommand{\Heoptb}{\hbox{{\rm He}\kern 0.1em{\sc i}}~$5876$}
\newcommand{\Heoptc}{\hbox{{\rm He}\kern 0.1em{\sc i}}~$4471$}
\newcommand{\Brgam}{\hbox{{\rm Br}\kern 0.1em$\gamma$}}
\newcommand{\Brten}{\hbox{{\rm Br}\kern 0.1em$10$}}
\newcommand{\Breleven}{\hbox{{\rm Br}\kern 0.1em$11$}}
\newcommand{\HeIh}{\hbox{{\rm He}\kern 0.1em{\sc i}}~$1.7$~{\micron}}
\newcommand{\HeIk}{\hbox{{\rm He}\kern 0.1em{\sc i}}~$2.06$~{\micron}}
\newcommand{\Teff}{\hbox{T$_{eff}$}}
\newcommand{\kms}{\hbox{km~s$^{-1}$}}

\newcommand{\cc}{\hbox{cm$^{-3}$}}
\newcommand{\Msol}{\hbox{M$_{\sun}$}}
\newcommand{\Mup}{\hbox{M$_{up}$}}
\newcommand{\peryr}{\hbox{yr$^{-1}$}}

\slugcomment{Paper status:  Accepted to the ApJ}

\shorttitle{Starburst Stellar Temperature Diagnostics}
\shortauthors{Rigby and Rieke}

\begin{document}


\title{Missing Massive Stars in Starbursts:  
Stellar Temperature Diagnostics and the IMF}


\author{J. R. Rigby and G. H. Rieke}
\affil{Steward Observatory, University of Arizona, 933 N. Cherry Ave., Tucson, AZ 85721}
\email{jrigby@as.arizona.edu,  grieke@as.arizona.edu}


\begin{abstract}
Determining the properties of starbursts requires spectral diagnostics
of their ultraviolet radiation fields, to test whether very massive stars
are present.  We test several such diagnostics, using new models of 
line ratio behavior combining Cloudy, Starburst99 and up-to-date 
spectral atlases \citep{pauldrach01,hillmill}.
For six galaxies we obtain new measurements of \HeIh/\Brten, a difficult 
to measure but physically simple (and therefore reliable) diagnostic.  
We obtain new measurements of \HeIk/\Brgam\ in five galaxies.
We find that \HeIk/\Brgam\ and [\OIII]/\Hbeta\ are generally unreliable 
diagnostics in starbursts.
The heteronuclear and homonuclear mid--infrared line ratios 
(notably [\NeIII]~$15.6$~\micron\ / [\NeII]~$12.8$~\micron) 
consistently agree with each other and with \HeIh/\Brten;
this argues that the mid--infrared line ratios are reliable diagnostics of
spectral hardness.
In a sample of $27$ starbursts, [\NeIII]/[\NeII] is significantly lower than 
model predictions for a Salpeter IMF extending to $100$~\Msol.  Plausible
model alterations strengthen this conclusion.
By contrast, the low--mass and low--metallicity galaxies II~Zw~40 and
NGC~5253 show relatively high neon line ratios, compatible with a 
Salpeter slope extending to at least $\sim40$--$60$~\Msol.
One solution for the low neon line ratios in the high--metallicity starbursts
would be that they are deficient in $\ga 40$~\Msol\ 
stars compared to a Salpeter IMF.  An alternative explanation, which we prefer,
is that massive stars in high--metallicity starbursts spend much 
of their lives embedded within ultra--compact \HII\ regions that 
prevent the near-- and mid--infrared nebular lines from forming and escaping.  
This hypothesis has important consequences for starburst 
modelling and interpretation.
\end{abstract}


\keywords{infrared: galaxies---galaxies: starburst---galaxies: 
stellar content---techniques: spectroscopic}

\section{INTRODUCTION}
\label{sec:intro}
In the very local ($D<5 h_{100}^{-1}$) Universe, the circumnuclear regions 
of just four galaxies (M82, NGC~253, NGC~4945, and M83) are responsible for
$\sim25\%$ of the current massive star formation \citep{heckman97}.  In these
``circumnuclear starburst galaxies'', the star formation is confined to the 
inner $0.2$ to $2$~kpc, in a dense, gas--rich disk where star formation rates
can reach $1000$ \Msol\ \peryr\ \citep{robhubble}. 
If the starburst initial mass function (IMF) includes
significant numbers of low--mass stars, then each starburst is
currently building up the stellar component of its host galaxy as well.
A starburst enriches and heats its interstellar medium, as well
as the local intergalactic medium. Starbursts can also 
drive large--scale winds that eject
interstellar gas, presumably casting metals into the voids and 
heating the gas between galaxies. Starburst galaxies thus 
play a number of important roles in galaxy evolution.   

If starbursts could be dated, then a sequence
could be pieced together, charting starburst evolution from 
triggering to post--starburst quiescence.  Starburst ages are most 
directly determined
by understanding the population of rapidly evolving massive stars.
The feedback effect of a starburst on its gas supply is 
transmitted through massive stellar winds and supernovae--driven superwinds.  
Thus, understanding the evolution of starbursts and their effects on the 
interstellar and intergalactic media both critically depend on understanding
the populations of massive stars.  

Unfortunately, since starburst galaxies are too far away to count individual 
stars, the high-mass IMF must be determined indirectly, in ways that are model--dependent
and crude.  \citet{leithconf} reviews these techniques and divides them
into three categories:  techniques to determine a lower mass cutoff by measuring 
the mass--to--light ratio; to find the slope of the IMF above $\sim10$~\Msol; 
and to determine an upper mass cutoff from the hardness of the ionizing
radiation field.

The ionizing spectrum is  
set by the starburst's age, IMF, and star formation history.
Consequently, the shape of the ionizing field spectrum is an important boundary 
condition on starburst models \citep{ghr-confproc}.  Ionizing continua
are often parameterized by an effective temperature (\Teff), as if
one stellar spectral type were responsible for the flux.
The UV spectrum cannot be measured directly because little ionizing continuum 
radiation escapes from a starburst \citep{lfhl}. Instead, the presence or 
absence of massive stars must be inferred using spectral diagnostics.  Extinction
in circumnuclear starbursts means that infrared diagnostics are preferred.

Many line ratios have been used to estimate starburst \Teff:
forbidden line ratios, mixed forbidden--recombination line ratios like 
$[\OIII]/\Hbeta$,  optical lines of \HeI, 
the near--infrared line \HeIk, and mid--infrared fine structure lines
for example.
Unfortunately, these diagnostics disagree by $2,000$ to $5,000$~K 
\citep{vr,thornley,ghr-confproc}, and suffer variously from intrinsic 
faintness, susceptibility to shocks and reddening, dependence on nebular 
conditions, and uncertain atomic constants.  

In this paper, we use the ratio of \HeIh\ to Brackett~10 (\Brten) to 
diagnose the hardness of starburst ionizing fields. The faintness of 
the \HeIh\ line restricts its measurement to nearby galaxies with strong 
emission lines.  In these galaxies, the \HeI/\Brten\ ratio 
should allow estimates of \Teff\ that are largely independent of reddening 
or nebular conditions.  We then use \HeI/\Brten\ to assess the accuracy of 
diagnostics that can reach distant galaxies.  Using diagnostics
we find reliable, we confirm that few massive starburst galaxies have
high--excitation spectra.  While this may occur because the IMF is biased
against high--mass stars, we propose that high--excitation spectra
are scarce because the massive stars spend most of their main sequence 
lifetimes embedded in ultracompact \HII\ regions. 

\section{OBSERVATIONS, DATA REDUCTION, AND CALIBRATION}
\label{sec:whatwedid}

To assist in evaluating T$_{eff}$ diagnostics, we have obtained new measurements 
of \HeIh/\Brten. This diagnostic is unaffected by metallicity, 
shocks, or level pumping. Regrettably, the \HeIh\ line is very weak, less than
$10\%$ the intensity of \Brgam.  Thus, the \HeIh/\Brten\ ratio can
only be measured in nearby starbursts with bright lines.   

For our sample, we chose six nearby starburst galaxies with 
large measured \Brgam\ fluxes and, when possible, supporting 
observations in the literature such as mid--infrared spectra. 
Near--infrared spectra were obtained on the nights of 2001 April 6 and 7, using
the FSPEC near--infrared spectrometer \citep{williams}
on the Steward Observatory Bok $2.3$~m telescope.

Table~\ref{tab:obs} lists target objects and integration times.
All observations were taken with the 
$600~$lines~mm$^{-1}$ grating, which produces effective resolutions of 
$R\approx 2000$ at $1.7$~{\micron} and $R\approx 3000$ at $2.1$~{\micron}.
The slit was $2.4$\arcsec\ by $90$\arcsec.  All exposures were guided 
by hand using an H--band camera that images the mirrored slit.  
Spectra of the six targets were obtained in H--band ($1.7$~\micron).
K--band ($2.1$\micron) spectra were also obtained unless high--quality 
spectra already existed in the literature.

The angular sizes of the nuclear starburst regions are small 
compared to the length of the FSPEC slit.  For each integration, 
nuclear spectra were obtained at four successive positions along
the slit.  For the calibration stars, six
spectra were taken along the slit.  (For brevity, we will call 
each resulting two-dimensional spectrum a ``frame'', and 
each group of frames in an integration a ``set''.)

\subsection{Data Reduction}
\label{sec:datared}
The infrared \HI\ and \HeI\ lines are faint, making the data reduction
approach critical. We therefore describe it in detail. The data were 
reduced using \texttt{iraf}.\footnote{IRAF is distributed by the National 
Optical Astronomy Observatories, which are operated by the Association of 
Universities for Research in Astronomy, Inc., under cooperative agreement 
with the National Science Foundation.}
First, dark frames were subtracted from the object frames.  Each frame was then 
flat--fielded using a median--averaged lamp flat.
The airglow and bias were removed by differencing neighboring
frames.  For the first and last frames of a set, the neighbor was
subtracted.  For each middle frame, the mean of the 
immediately--bracketing frames was subtracted.  For the calibration 
stars the subtraction was simple.
Over the longer integration times required for the galaxies 
($\sim4$ minutes per frame), the sky background is variable.  
Accordingly, prior to subtraction, 
we scaled each galaxy frame by a constant, generally within a percent 
of unity, to optimize the sky cancellation.
This technique of differencing neighboring two--dimensional spectra
usually removes the sky emission lines accurately.
For a few galaxies, the resultant sky subtraction was
not adequate.  In these exposures, the sky lines in a set of 
frames were offset along the dispersion axis by $0.005$ to $0.01$~pixels,
suggesting a slight, monotonic shift in the grating tilt.  
To improve the cancellation, we used
\texttt{onedspec.identify} and \texttt{onedspec.reidentify} to fit, 
for each frame, a linear shift in the position of the airglow lines 
with respect to a reference frame.   We then used 
\texttt{images.imgeom.imshift} to shift 
the frames, using linear interpolation, to zero the offsets.
This was done for H--band frames of NGC~4861, He~2--10, NGC~4102,
and NGC~3504 as needed.

The next step in the reduction was to combine the frames within a
set.  Offsets were determined by summing each frame down the dispersion
axis and measuring the location of the continuum peak in the resulting
one-dimensional image.
Frames were magnified by a factor of six to permit fractional pixel shifts,
which minimizes smearing of the data and maximizes preservation of flux
during the next step of straightening.  
Magnified images were then remapped to make the spatial axis perpendicular 
to the dispersion axis.  This remapping is accomplished by fitting the
tilt of the dispersion axis as a cubic polynomial; this function is well--defined 
and does not vary with time.  Straightening the spectra in this manner 
simplifies subsequent extraction, stacking, and wavelength calibration
\citep{chad_reduction}. 

Frames were then median combined with \texttt{imcombine} using no 
rejection, and using scale and weight factors determined from each 
frame's median continuum strength (generally within $20\%$ of unity).  
Frames were then de-magnified.
Bad pixels were replaced with the average of their immediate two 
neighbors along the dispersion axis.  

A similar procedure created two--dimensional sky frames suitable
for wavelength calibrations.  The frames were straightened and
median combined as before, but without weights, scales, or offsets.
The resulting images have no continua, only sky lines.  A high
signal--to--noise 1--D sky spectrum was then created by taking, at each
wavelength, the median value over all spatial positions.  

We then extracted the spectra of the galaxies and calibration 
stars.  Using \texttt{apall} in \texttt{iraf}, we traced each continuum 
interactively using low--order Legendre polynomials.  The aperture
width was chosen to be where the signal at the brightest part of
the continuum dropped to $30\%$ of peak.  No extra background was
subtracted at this point; trials with additional subtraction (to
remove residual sky lines) added more noise than they eliminated.
Spectra were extracted from the sky frames by using the apertures
fitted for the corresponding objects.

Next, the spectra were corrected for telluric absorption.  To do 
this, each galaxy spectrum was first divided, in pixel 
space, by the spectrum of a calibration star.  
When a target observation was bracketed by calibration star 
observations, the average of the two stellar spectra was 
divided into the target spectrum.  Otherwise, the calibration 
spectrum taken closest in time to the target was used.
Calibration stars were dwarfs of spectral types F6 to G0, most 
within $5\degr$ of the target object.  Such stars have relatively
featureless intrinsic spectra at $\sim2$~{\micron}, so their observed 
spectra reflect the variable absorption of the Earth's atmosphere.

We wavelength--calibrated
the sky spectra of the galaxies, using \texttt{onedspec.identify}
and tabulated vacuum wavelengths of the OH lines calculated by 
C.~Kulesa (1996, private communication).  Scatter in the wavelength
calibration of the galaxies was less than $\pm0.2$~{\AA}, and
usually below $\pm0.1$~{\AA}.

Wavelength solutions of the sky spectra were transferred to the
[target/calibrator] spectra.  
Due to the longer integration times, the sky spectra derived 
from galaxy frames give wavelength solutions more precise 
than those derived from sky spectra of the star frames.  
The wavelength solution is quite stable with time.

The calibration stars are not completely featureless; their
metal absorption lines produce spurious emission lines in the 
[target/calibrator] spectra.  To correct for this effect, we multiplied
the [target/calibrator] spectra by the solar spectrum, which has
been corrected for atmospheric absorption \citep{lwsun}.  The solar
spectrum was first normalized and broadened to $10$~{\AA} in H 
and $9$~{\AA} in K to match our observations.  Since the calibration
stars are similar in spectral type to the Sun, the solar multiplication
cures the final spectrum of the metal lines and the global 
Rayleigh--Jeans slope that the calibration star introduced.
The resulting spectrum is [target/calibrator]$*$[$\sun$].
This procedure is explained more fully by Maiolino, Rieke, \& Rieke (1996).
\nocite{mrr}

At this point, NGC~4861 and NGC~4214 still showed residual sky lines.
The spectra were improved by subtracting a scaled, extracted sky spectrum.
For both galaxies, \Breleven\ and \HeIh\ are uncontaminated by OH lines,
whereas \Brten\ is somewhat contaminated in NGC~4861 and seriously 
contaminated in NGC~4214 (the lowest redshift galaxy in our sample).  
Accordingly, we will consider only \Breleven\ in lieu of both \Brten\ and
\Breleven\ for these two galaxies. 

\subsection{Combining Spectra}
\label{sec:combspec}
For the H--band spectra, we observed targets for $0.5$~hr between
calibrators.  Because our total integration times on each galaxy 
were substantially longer than this, the spectra must be combined.  
To do this, we used \texttt{onedspec.scombine}, median combining 
groups of $\ge4$~images, and average combining otherwise.

The \HeIk\ and \Brgam\ lines were covered by different grating settings.  
We spliced together the two grating settings for each target by scaling
the $2.085$~{\micron} spectrum by a constant until it matched the flux 
level of the $2.15$~{\micron} spectrum in the region of overlap.  
Scaling constants were between $1.08$ and $1.3$.  We then joined the
spectra using \texttt{scombine}.

\subsection{Subtracting the Stellar Continuum}
\label{sec:contsub}
The nebular lines we seek sit atop a stellar continuum, whose absorption
lines can mask or alter the emission line ratios.  Before faint emission line
fluxes can be measured, the stellar continuum must be removed.

The lines of interest at K (\Brgam\ and \HeIk) are strong enough that
subtraction of the stellar continuum is not necessary.  For H--band, 
we used a stellar continuum template made by combining the $15$ stars 
from a stellar atlas (observed with the same spectrometer: 
V.~D.~Ivanov, in preparation) which 
minimized the residuals when subtracted from NGC~$253$ \citep{chad}.
Thus, the template was chosen to be a good fit to a $\sim$solar metallicity 
starburst galaxy, and was not made by modelling the stellar 
populations of each galaxy in our sample.   The stars in the template
are of stellar types K0 to M3, with metallicities between solar
and half--solar.  Two of the stars are supergiants,
five are bright giants, five are giants, and three are dwarfs.

The resolution of the template is $130$~{\kms} in H.  To match 
the intrinsic velocity dispersion of the galaxy spectra, we convolved 
this template with a Gaussian
kernel as necessary to lower the resolution.  He~2--10 and NGC~3077
were best fit with no convolution; NGC~3504 required a template with 
$150$~{\kms} resolution, and NGC~4102 required a $175$~{\kms} template.
For NGC~4214 and NGC~4861, continuum subtraction was unimportant because
the continua are much weaker than the nebular emission lines.

The stellar continuum template has an absorption feature at 
$1.7010$~{\micron}.  By comparison with spectra of the Sun \citep{lwsun} 
and Arcturus \citep{mont}, we identify this feature as a blend of three 
nickel lines and four (weaker) iron lines.  At a resolution of 
$130$~{\kms}, the blend has an equivalent width of 
$\le 1.5$~{\AA}, and dips to $94\%$ of the continuum level.
Because the absorption feature is a metal line blend, its 
strength will depend on metallicity.  

Besides a simple subtraction of the stellar continuum, we also 
added a $20\%$ featureless continuum to the template, renormalized,
and subtracted the new continuum from the galaxy spectra.  This 
procedure crudely approximates the effect of lower metallicity. These two 
realizations of the continuum subtraction provide some estimate of 
the associated uncertainty.

\begin{figure*}
\figurenum{1}
\plotone{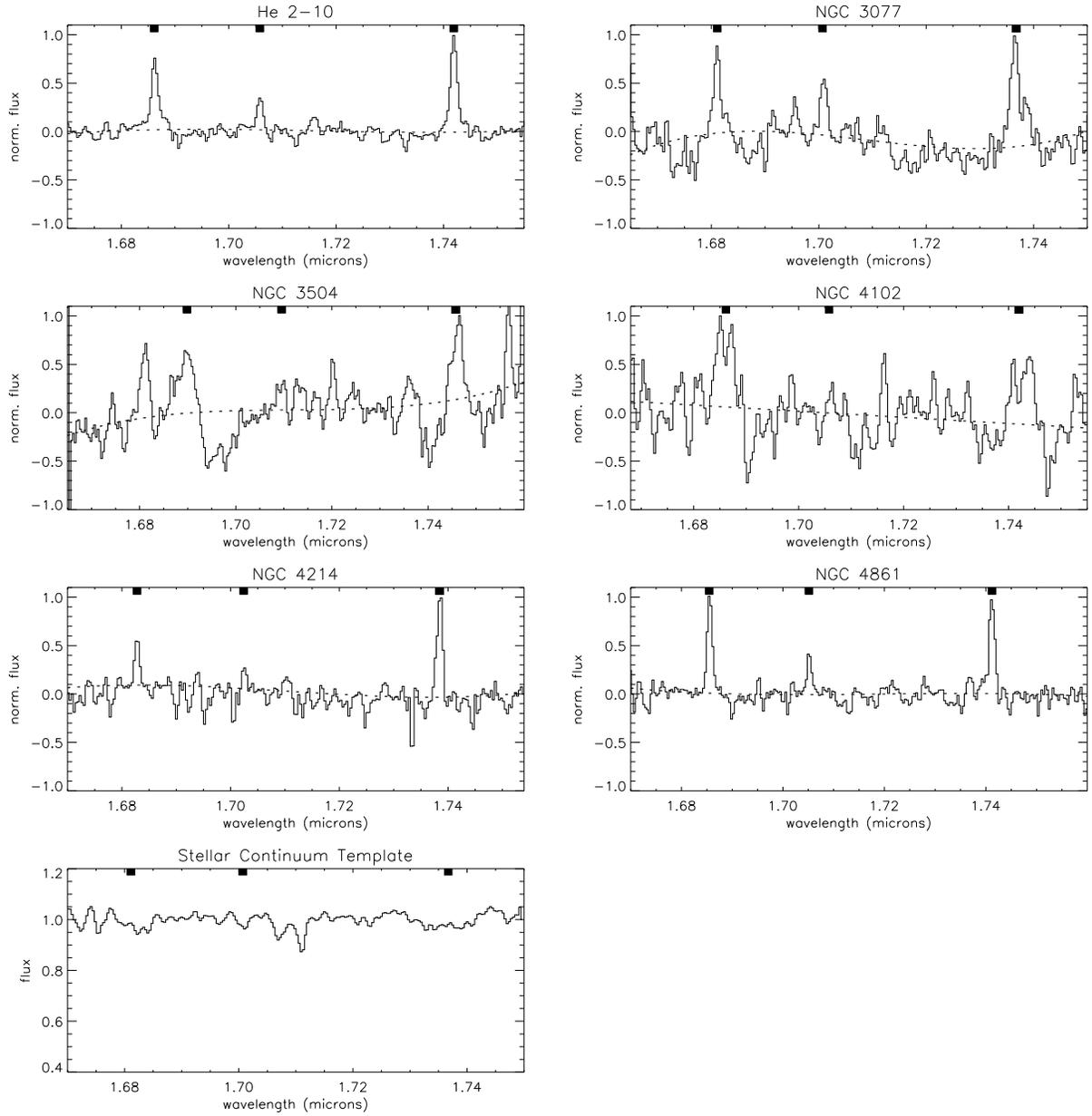}
\figcaption{H--band spectra of the starburst galaxies in our sample.
For presentation,
the median continuum levels were set to zero, and the spectra normalized so
that the maximum values were unity.  The galaxy spectra are plotted versus 
observed (redshifted) wavelength.  The expected positions of 
\Breleven, \HeIh, and \Brten\ are marked.  Dashed lines show the continuum
fits.  
(Note that \Brten\ in NGC~4214 is contaminated by a telluric OH line.)  
For reference, the rest--frame 
stellar continuum template is also plotted, without renormalization.}
\label{fig:spec_17}
\end{figure*}

\begin{figure*}
\figurenum{2}
\plotone{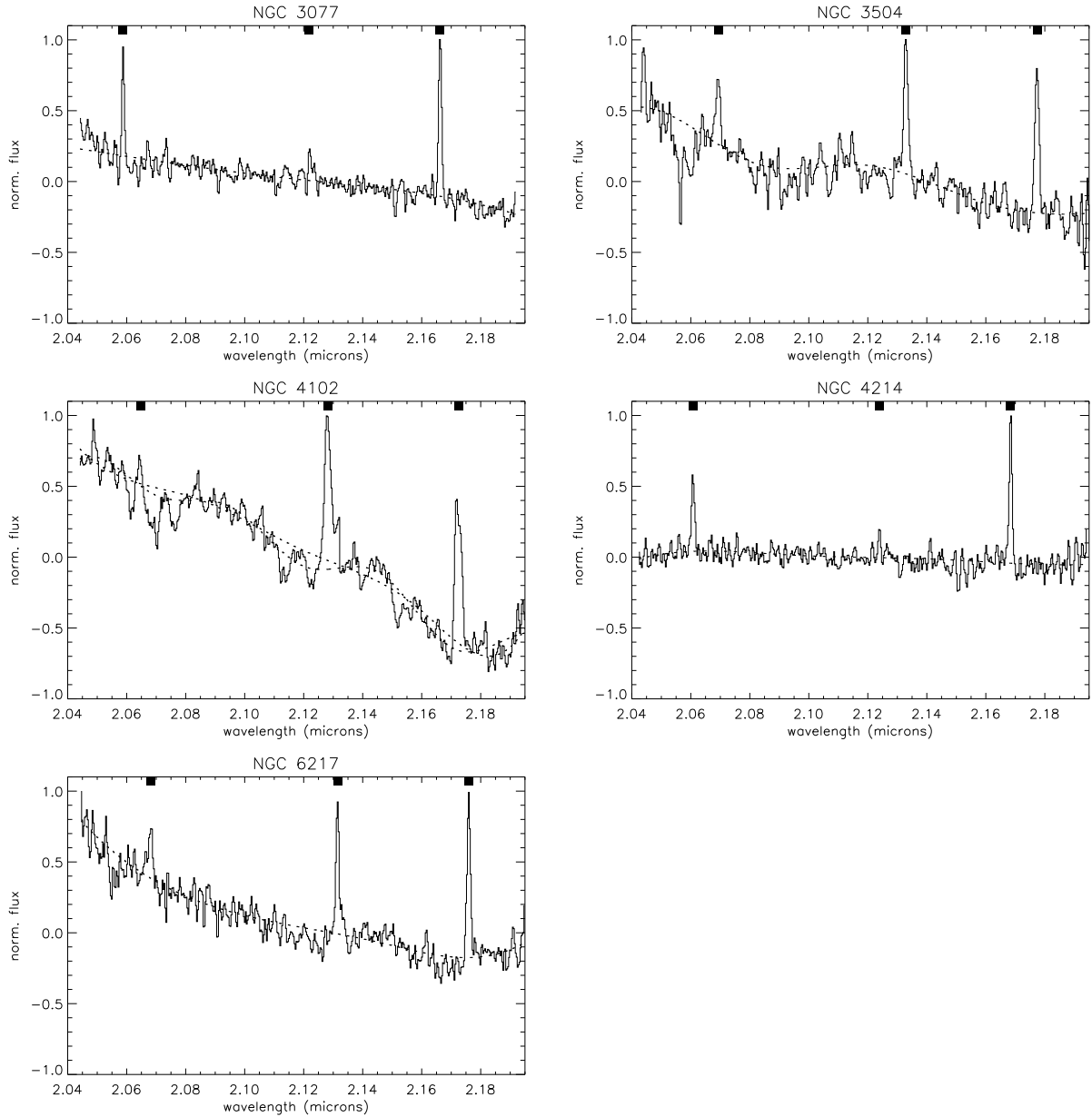}
\figcaption{K--band spectra of the starburst galaxies in our sample. 
Spectra are normalized as in figure~\ref{fig:spec_17}, and wavelengths 
are observed.  The expected positions of \HeIk, H$_2$~$2.12$~{\micron}, 
and \Brgam\ are marked.  Dashed lines show continuum fits. 
For NGC~4102, two different continuum fits were used. 
}
\label{fig:spec_20}
\end{figure*}

Continuum--subtracted H--band spectra of the galaxies and the stellar 
template are plotted in figure~\ref{fig:spec_17}.  
K--band spectra are plotted in figure~\ref{fig:spec_20}.

\subsection{Measuring Line Ratios}
\label{sec:lineratios}

To set the continuum level, we fit a low--order Chebyshev function 
across each spectrum, excluding emission lines from the fit.
Each line was fit by a Gaussian to measure the line fluxes listed
in table~\ref{tab:waves}.
For the noisy, non--Gaussian line profiles of NGC~$4102$, 
we directly summed flux rather than fit Gaussians.

To measure the relative strength of \HeIh, we considered two ratios:  
\HeIh/\Brten, and $[\HeIh / \Breleven]\times[\Breleven / \Brten]_{case B}$.
We assumed the value  $[\Breleven / \Brten]_{case B} = 0.75$, which is 
appropriate for $n = 10^2$~{\cc} and $T_e = 5,000$~K \citep{chad,hs87}.
For each galaxy, we computed both these ratios for both realizations of
the continuum subtraction (with or without the $20\%$ featureless continuum), 
and used the mean of these four values as the \HeIh/\Brten\ ratio,
and the standard deviation as an estimate of the uncertainty associated 
with the continuum subtraction.  We also computed the ratio \HeIk/\Brgam. 
Our measured line ratios are listed in table~\ref{tab-line-rats}, along
with values of \HeIk/\Brgam\ from the literature, and a weighted mean for 
\HeIk/\Brgam\ that combines new and literature values.  (The quoted
uncertainty for the weighted mean is the error in the mean.)

Because of the $1.7$~\micron\ stellar absorption feature, studies that do not
subtract the continuum in galaxies with weak \HeIh\ will somewhat underestimate 
the \HeIh\ line strength and therefore underestimate \Teff.  To test the
magnitude of this effect, in table~\ref{tab-line-rats} we list both 
continuum--subtracted and raw (un--subtracted) \HeIh/\Brten\ ratios.
For galaxies NGC~3504 and NGC~4102, the lines are so weak relative to
the stellar continuum that the \HeIh/\Brten\ cannot be measured without
continuum subtraction.  For the other galaxies in table~\ref{tab-line-rats}, 
the raw and continuum--subtracted line ratios are very similar; for these
galaxies (He~2--10, NGC~3077, NGC~4214, and NGC~4861), the continuum
subtraction is not an important source of uncertainty.

%
\section{MODELLING LINE RATIO BEHAVIOR}
\label{sec:models}
Because the emission lines used to diagnose effective temperature have 
different excitation energies, one cannot verify that a particular
diagnostic works by simply testing whether it exhibits a one-to-one
correlation with another diagnostic.  Instead, one must test diagnostics in 
light of photoionization models that, given realistic hot stellar ionizing 
spectra, predict line ratios appropriate to idealized nebulae.
One can then ask a) whether the observed ratios populate the 
line ratio space permitted by models; and b) whether 
many observed line ratios for a particular galaxy are consistent, that is, 
can all be produced by one set of physical parameters.  Thus, 
translating a nebular line ratio to a statement
about stellar content is necessarily model--dependent.  

Past studies have run  series of models in which a single main sequence star 
photoionizes a nebula, producing tabulated line ratios as a function of stellar \Teff.  
A measured galactic line ratio is then translated into an effective temperature 
using this tabulation \citep{dpj,al,vr,bkl,fs-m82}.
This method has its uses:  namely, to compare model inputs and assumptions, 
and understand what line ratios different stellar classes can produce.  
However, in \S~\ref{sec:specsynth} we will argue that, especially 
for mid--infrared line ratios and \HeIh/\Brten, starbursts are poorly 
approximated by single main sequence stars;  
to translate a line ratio into a meaningful statement about a stellar 
population, one must consider the flux from \emph{all} the stars as a 
function of time.  First, though, we will consider the insights and limitations 
of simple one--star models.

All our models (single--star and population synthesis) are radiation--bounded 
thin shells created by the photoionization code Cloudy~94.00 \citep{hazy}.  
Using parameters determined by \citet{fs-m82} for M82, we set 
the total hydrogen number density to $n_H = 300$~\cc\ and inner radius to 
$R=25$~pc.  This choice of radius produces line ratios within $2\%$ of the 
plane--parallel ($R=\infty$) case.  This is because the shell is thin. 
Thus, the choice of radius only slightly affects the models, which are
effectively plane--parallel.
For single--star models, we use a constant ionization parameter of 
$\log U = -2.3$.  
We ran two sets of models, one with gas--phase abundances of solar, and the other
with $1/5$ solar abundances (``the low--Z models''); neither abundance set includes 
depletion onto dust grains.  (We address the effects of dust in \S~\ref{sec:caveats}.)
Because Cloudy does not predict the intensity of \HeIh, we scaled the
intensity from \Heoptc~\AA, which shares the same upper level.
For Case B and T$_e=5,000$~K, the \HeIh\ line is a factor of $7.4 \times 10^{-3}$ 
fainter than \Heoptc~\AA.

\subsection{Models Using Individual Stars}
\label{sec:indystars}
We first consider the ratio of [\NeIII]~$15.6$~\micron\ to 
[\NeII]~$12.8$~\micron.  For reference, it requires $22$~eV to make
singly--ionized neon, and $41$~eV to make doubly--ionized neon.
We took ionizing spectra from the O~star models of Pauldrach, Hoffmann, \& Lennon 
(2001), as prepared by \citet{snc},  
\nocite{pauldrach01}
and also the CoStar model spectra of \citet{sk} as hardwired in Cloudy.  
These two stellar libraries predict dramatically different line ratios.  
Dwarf, giant, and supergiant
Pauldrach stars all produce a maximum [\NeIII]/[\NeII] ratio of $10$ at \Teff$=50,000$~K.  
By contrast, the CoStar dwarf and giant atmospheres yield [\NeIII]/[\NeII]$=40$ at
\Teff$=50,000$~K.  At \Teff$=35,000$~K, the predicted line ratios disagree by an 
order of magnitude. \citet{fs-m82} used Pauldrach atmospheres and an earlier 
version of Cloudy to make their figure~8, which our Pauldrach models reproduce.

The other mid--infrared line ratios also show this discrepancy.
CoStar models predict ten times higher [\SIV]/[\SIII] and [\SIV]/[\NeII] ratios
than Pauldrach models for most of the $25,000<\Teff<50,000$~K range; for [\ArIII]/[\ArII]
and [\ArIII]/[\NeII], CoStar gives $2$ and $3$ times higher ratios, respectively.  The
near--infrared line ratios \HeIh/\Brten\ and \HeIk/\Brgam\ are not sensitive to the 
choice of stellar atlas.

It is sobering that current O~star models predict such different mid--infrared 
line ratio strengths.  On the bright side, this sensitivity suggests that mid--IR 
line ratios may provide astrophysical tests of O~star spectral models in simple 
\HII\ regions.  \citet{giveon} performed such a test in Galactic \HII\ regions; 
they find that \citet{pauldrach01} atmospheres fit the observed [\ArIII]/[\ArII] 
versus [\NeIII]/[\NeII] relation, whereas stellar models that assume LTE do not.  
In another test, \citet{pauldrach01} argued that their models successfully 
reproduce the  observed far--UV spectra of hot stars, as opposed to other models.
Finally, \citet{snc} argue that \citet{pauldrach01} atmospheres should be 
more realistic than those of  \citet{sk} because the latter neglect 
line broadening, and thus underestimate line blanketing.  As a result, CoStar 
atmospheres have significantly higher ionizing fluxes, especially at energies
exceeding the He$^+$ edge.

Given these problems,
and that the CoStar atmospheres predict much higher [\NeIII]/[\NeII] line ratios than are 
observed, we will use the \citet{pauldrach01} atmospheres in this paper.  Still, that the
Pauldrach spectra are better does not mean they are correct; the CoStar--Pauldrach
discrepancy should serve as some warning of the current uncertainties regarding hot 
star spectra---a critical input to the models.

We also consider Wolf--Rayet (WR) stars in simple nebulae.  We use WN and WC model 
spectra compiled by \citet{snc}, which were generated using the code of \citet{hillmill}.
For a given \Teff, these model WR stars yield much lower [\NeIII]/[\NeII] 
ratios compared to Pauldrach O~stars:  at $\Teff=50,000$~K, the difference is a factor 
of $30$ for WN, and a factor of $10^4$ for WC stars.  This is because of the
very strong line blanketing found in WR stars.  
Since Wolf--Rayet stars can reach much hotter temperatures
than main sequence stars, a $\Teff \la 140,000$~K WC star can reach $[\NeIII]/[\NeII]\sim10$
(comparable to the ratio produced by a $\Teff=50,000$~K Pauldrach O~star).  Similarly,
a $120,000$~K WN star can reach [\NeIII]/[\NeII]$=100$.  Thus, given these stellar 
atmospheres, only a WN star can give rise to a neon ratio between $10$ and $100$.

As a result, there are mid--infrared line ratio regimes that only Wolf--Rayet stars 
can populate (again assuming solar metallicity.)  For maximum effective temperatures 
of \Teff$^{MS}=50,000$~K, \Teff$^{WN}=120,000$~K, and \Teff$^{WC}=150,000$~K,
for \cite{pauldrach01} model O~stars and \citet{hillmill} WR stars, we find the following:
\begin{itemize}
\item  Main sequence O~stars can only produce $[\ArIII]/[\ArII] \le 18$, whereas 
       WN and WC stars can reach ratios of $40$.
\item  MS O~stars can only produce [\SIV]/[\SIII]$=0.5$, while WC stars can reach $1.2$, 
       and WN can reach $3$.
\item  MS O~stars can only produce [\SIV]/[\NeII]$=4$, whereas WC can reach $12$ and 
       WN can reach $130$.
\item  MS O~stars and WC stars can only produce [\ArIII]/[\NeII]$=2.5$, 
       whereas WN stars can reach $14$.
\end{itemize}

We have just seen that the conversion from mid--infrared line ratio to \Teff\ 
is very different
for main sequence stars than for Wolf--Rayet stars.  Thus, even a modest portion 
of WR stars within a hot stellar population can significantly affect the line ratios.  
We also conclude that in
a solar--metallicity starburst, if the line flux ratios exceed the maximum that 
main sequence stars can produce, then WR stars dominate the ionizing flux.  

\subsection{Spectral Synthesis Models}
\label{sec:specsynth}
Given the influence of WR stars, we must consider the more realistic scenario 
of an evolving stellar population as the ionizing source.  We used the 
spectral synthesis code Starburst99 
version~4.0 \citep{starburst99} to create instantaneous starbursts 
with an initial mass function of Salpeter--slope \citep{salpeter} and
initial stellar masses between $1$~M$_{\sun}$ 
and a variable upper mass cutoff, ``\Mup'' 
(\Mup$=100$, $75$, $60$, $50$, $40$, and 
$30$~\Msol.) (\Mup\ in this paper always refers to the IMF, not 
the present-day mass function.)  As in our single--star models, 
this version of Starburst99 uses O~star model spectra from \citet{pauldrach01} 
and Wolf--Rayet model spectra from the code of \citet{hillmill}, as prepared by 
\citet{snc}.  

We created two suites of models:  the first set assumed solar metallicity
in Starburst99 and Cloudy, and the ``solar metallicity, high--mass loss'' option, 
which is recommended as the default for Starburst99.  
(The alternative ``standard mass loss'' option gives qualitatively similar results.)
The second set of models used a gas--phase metallicity of $1/5$~solar in Cloudy,
and the ``high--mass loss, Z=$1/5$~solar'' and ``uvlines = Magellanic'' settings in
Starburst99.  Dust was ignored (and will be addressed in \S~\ref{sec:caveats}.)  Other
parameters were set as for the single--star models.
Starburst99 calculated the spectral energy distribution (SED) of the burst 
every $0.1$~Myr for $10$~Myr after the starburst.    
The ionization parameter was normalized to a maximum value of $\log U = -2.3$,
and scaled by the number of hydrogen--ionizing photons present in the SED.
Given the SEDs as input, Cloudy calculated line ratios as a function of starburst age.

\begin{figure*}
\figurenum{3}
\epsfig{figure=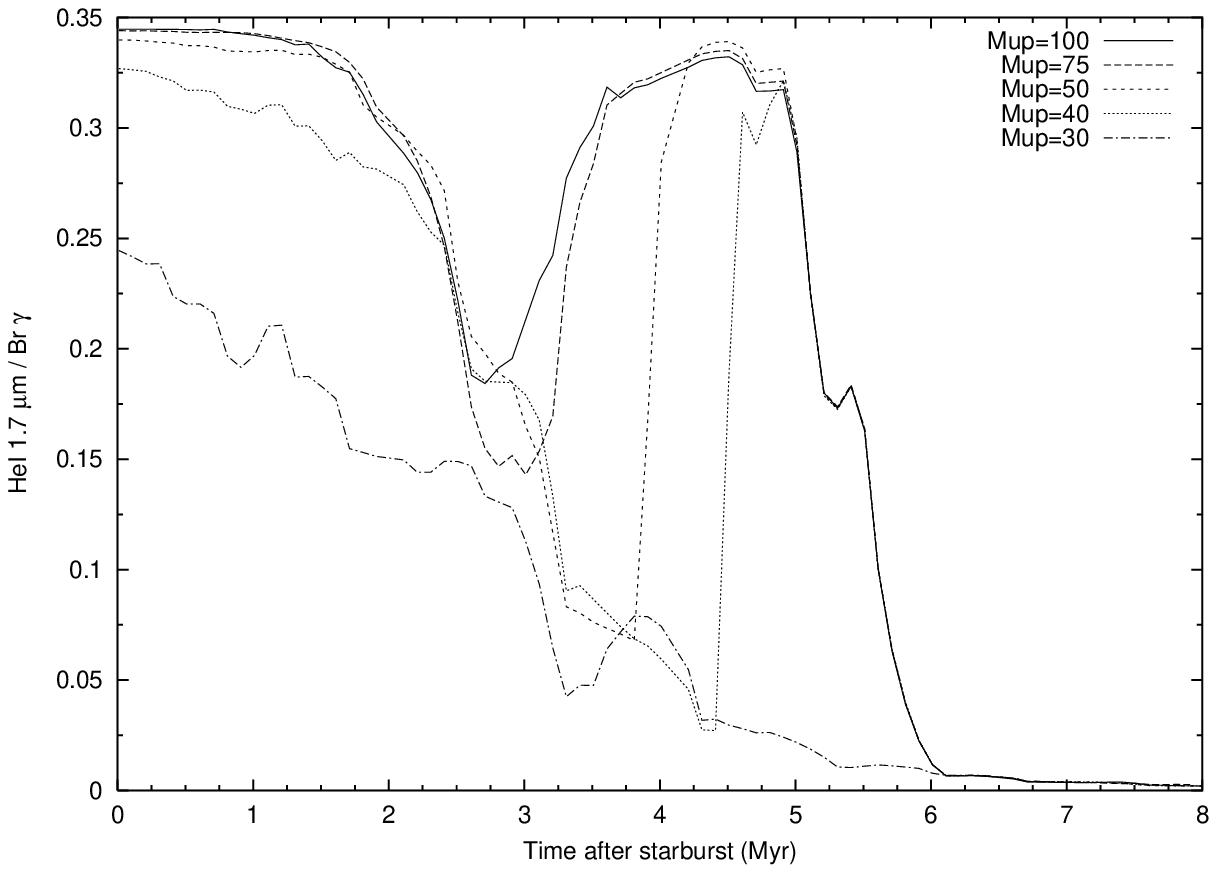,height=2in,angle=0,clip=t}
\epsfig{figure=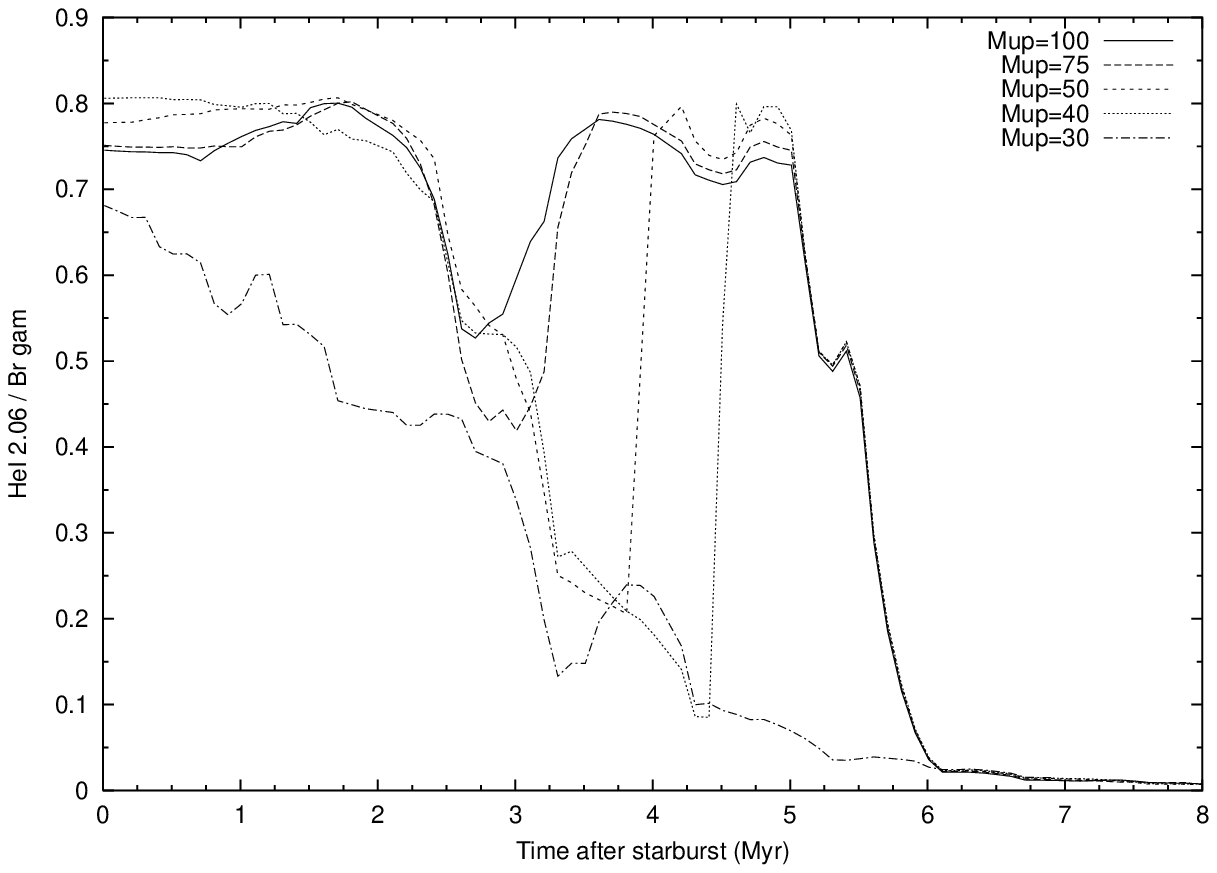,height=2in,angle=0,clip=t}
\epsfig{figure=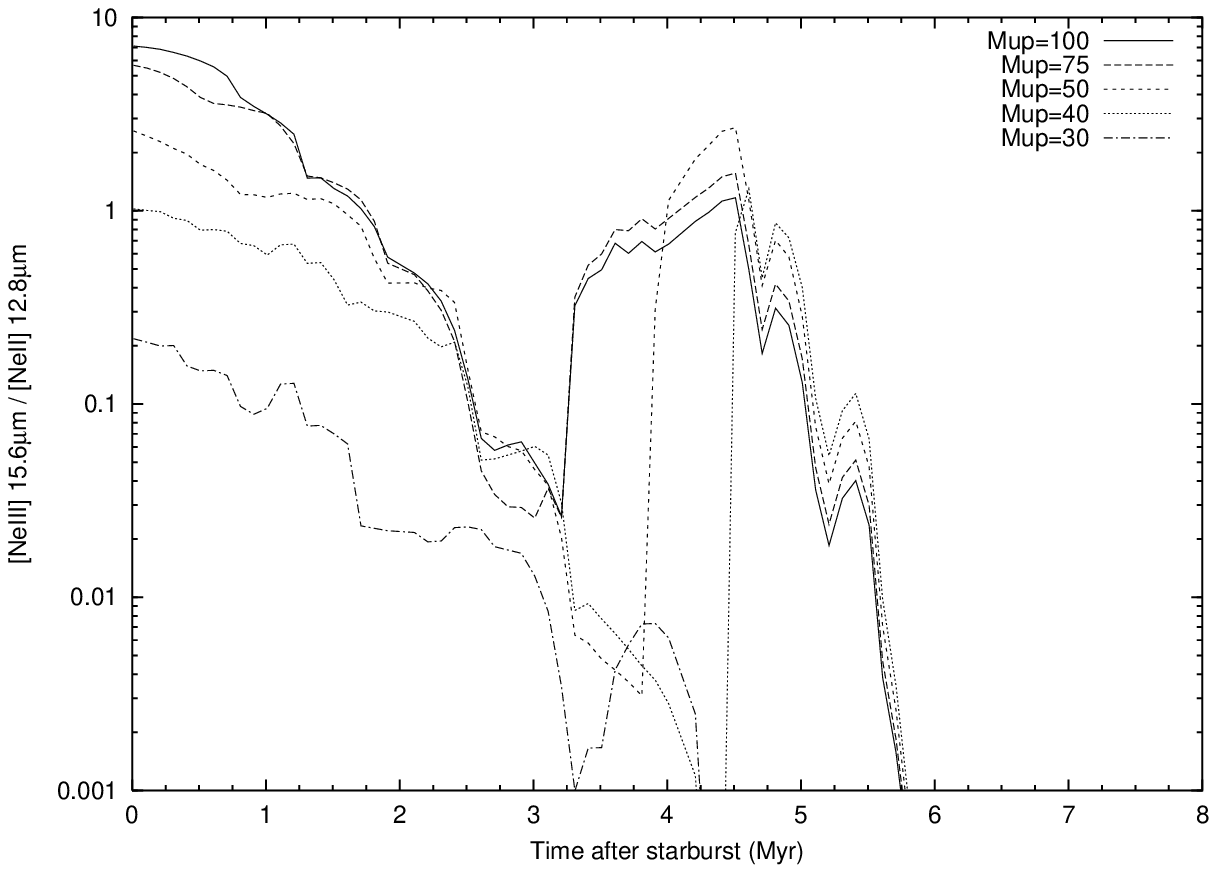,height=2in,angle=0,clip=t}
\epsfig{figure=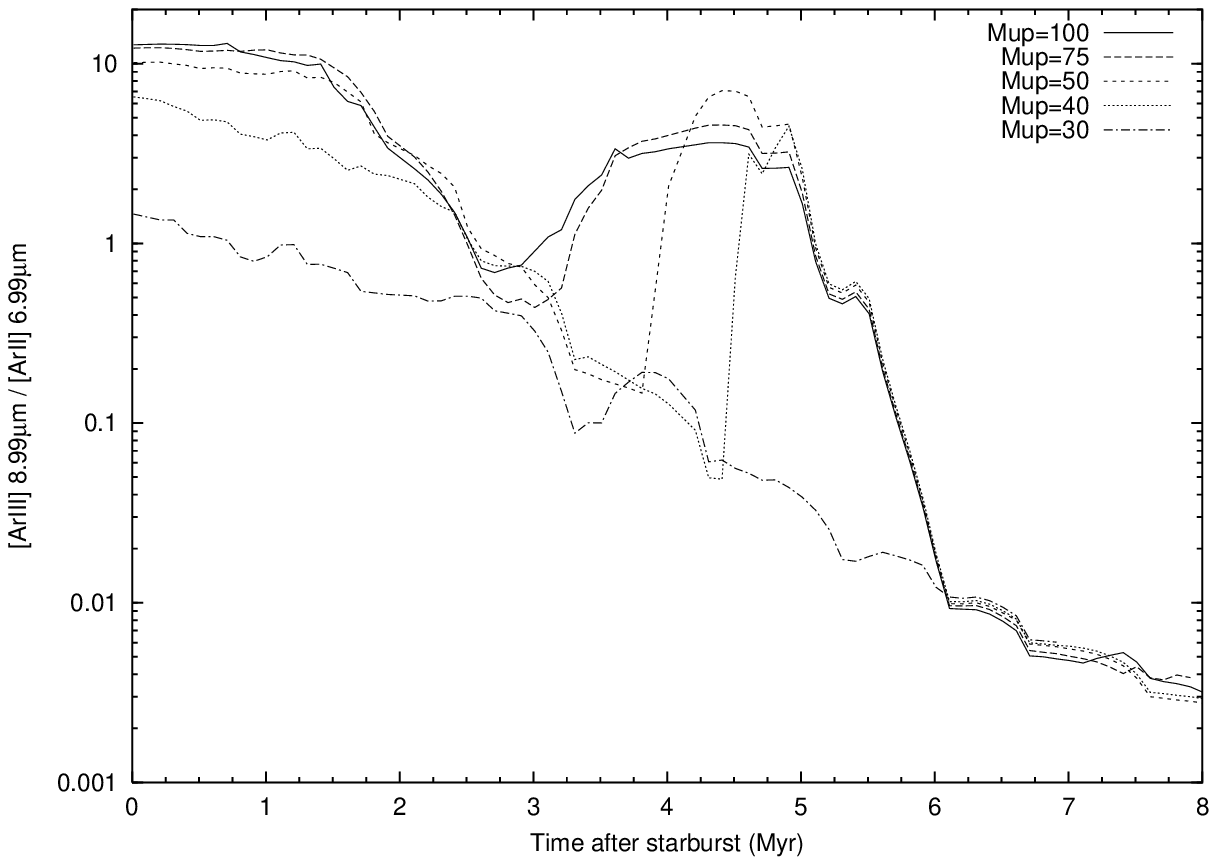,height=2in,angle=0,clip=t}
\epsfig{figure=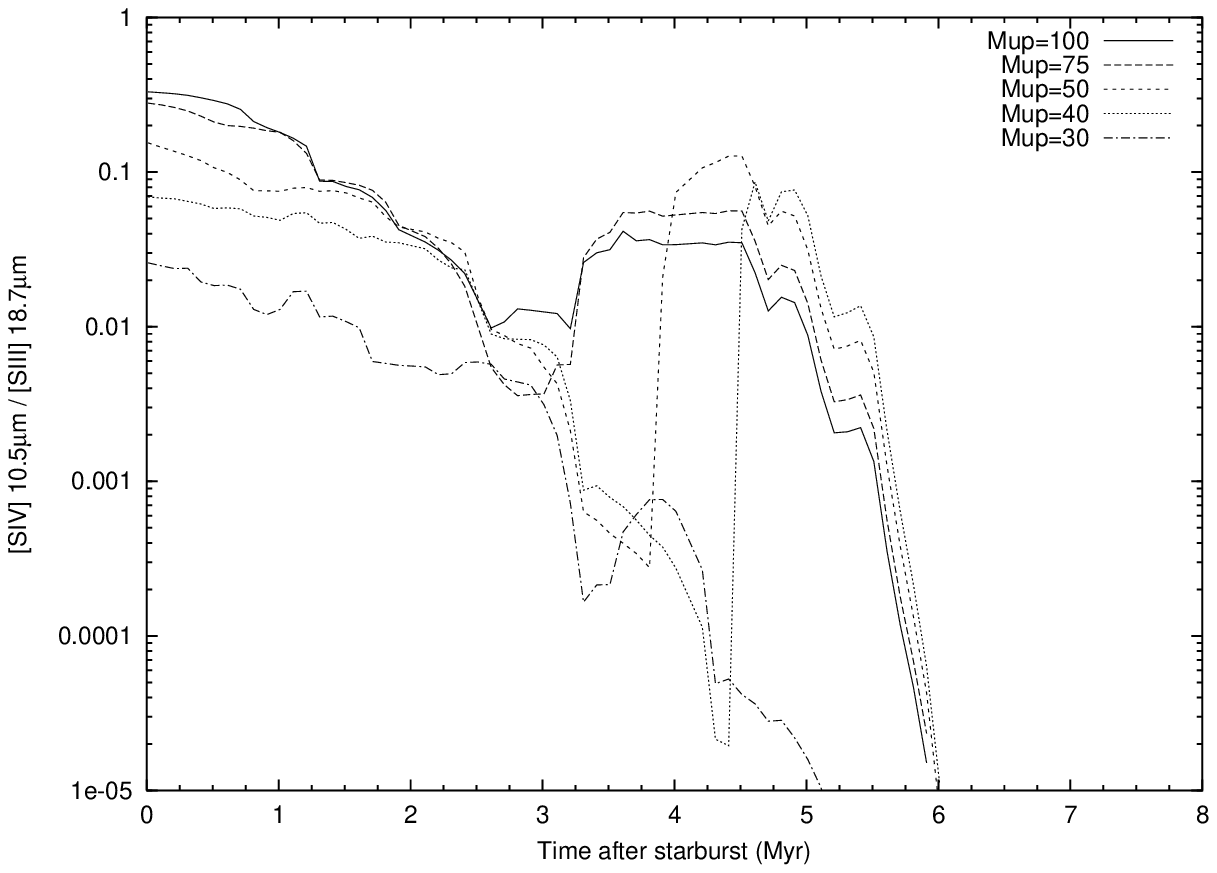,height=2in,angle=0,clip=t}
\epsfig{figure=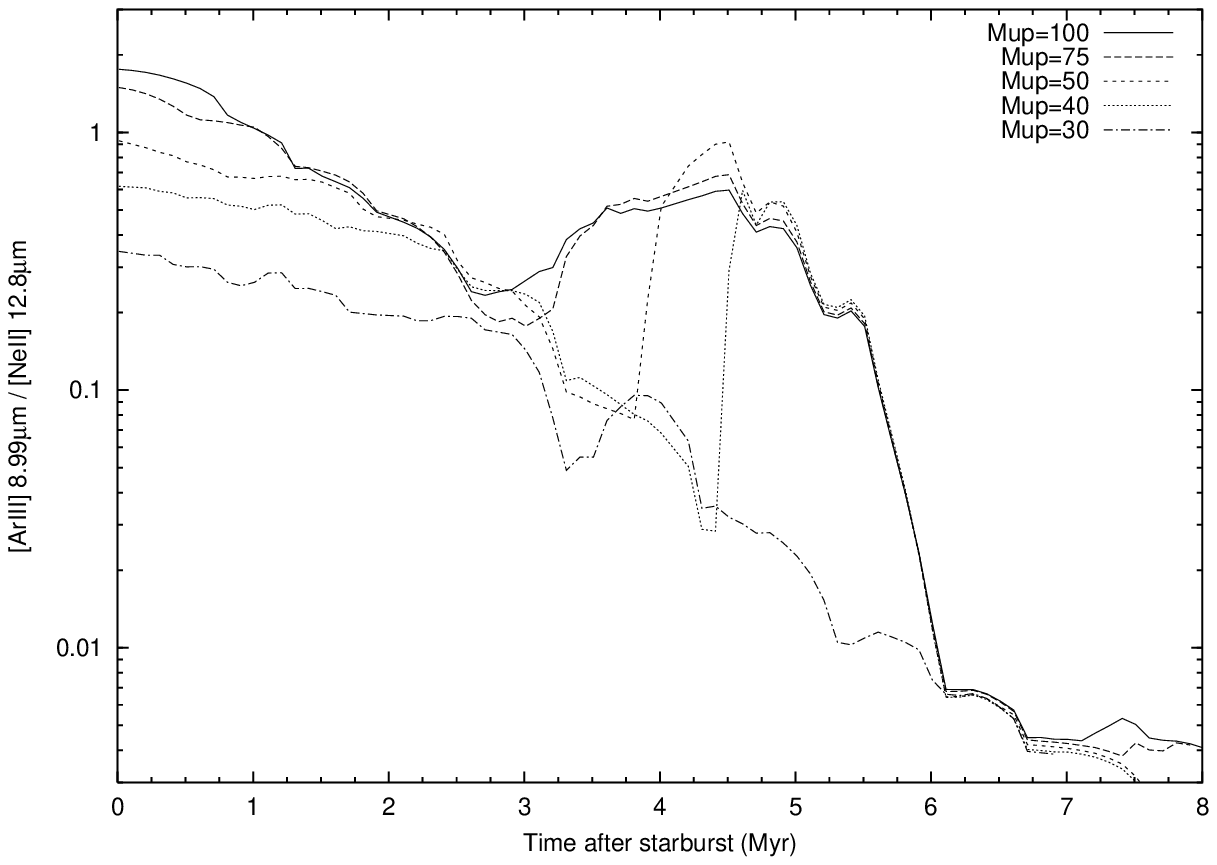,height=2in,angle=0,clip=t}
\epsfig{figure=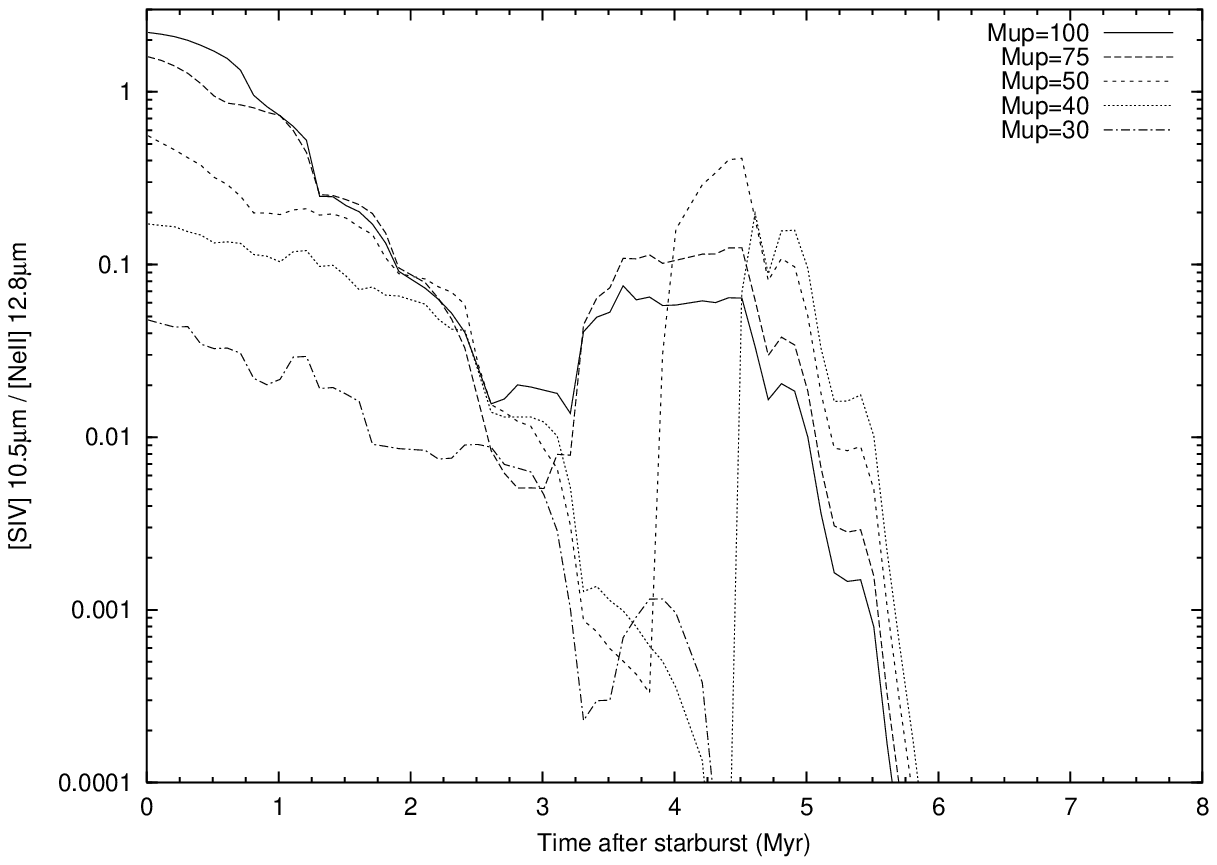,height=2in,angle=0,clip=t}
\figcaption{Line ratios versus time for Cloudy photoionization models with 
Starburst99 SEDs.  Models have solar metallicity, and 
assume an instantaneous burst of star formation and an 
IMF with Salpeter slope and stars with masses between $1$~\Msol\ and an
upper mass cutoff (\Mup).
}
\label{fig:models}
\end{figure*}

\begin{figure*}
\figurenum{4}
\epsfig{figure=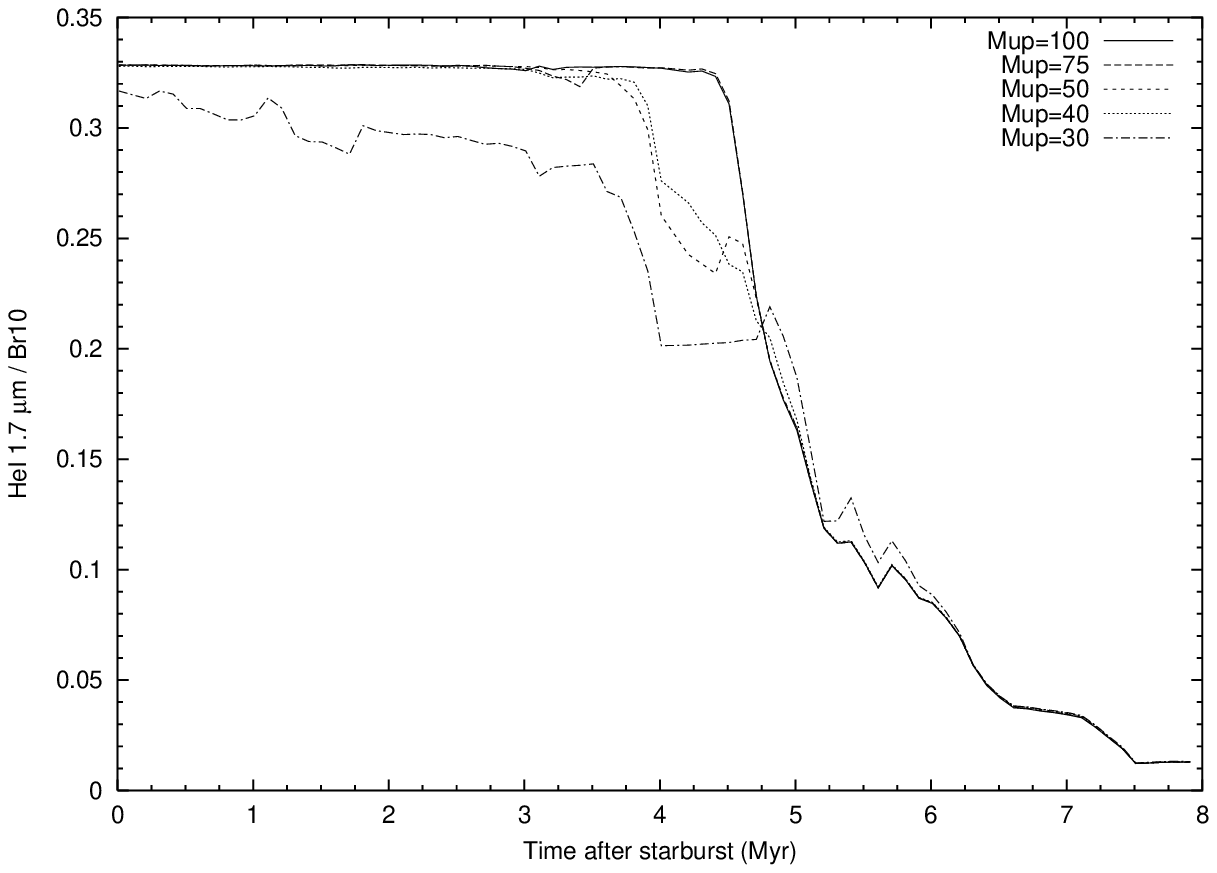,height=2in,angle=0,clip=t}
\epsfig{figure=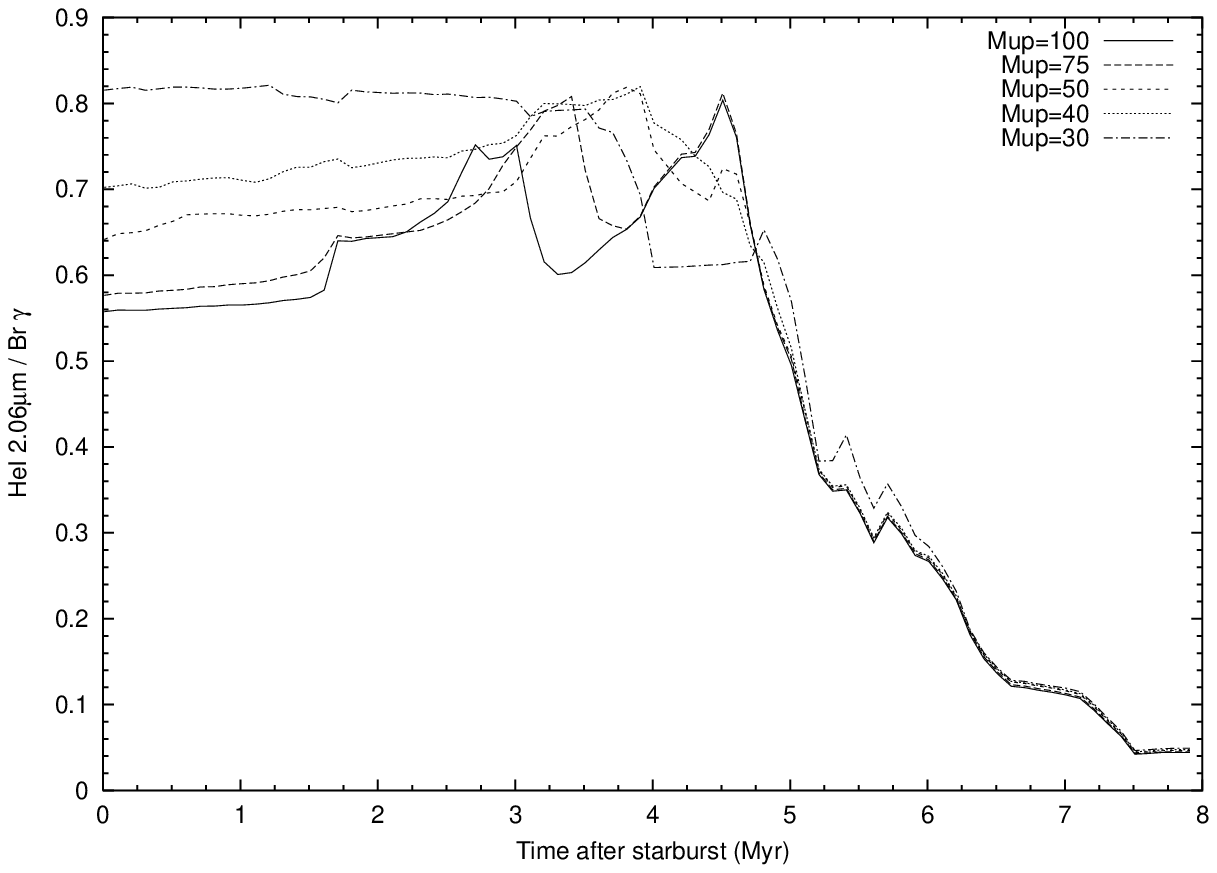,height=2in,angle=0,clip=t}
\epsfig{figure=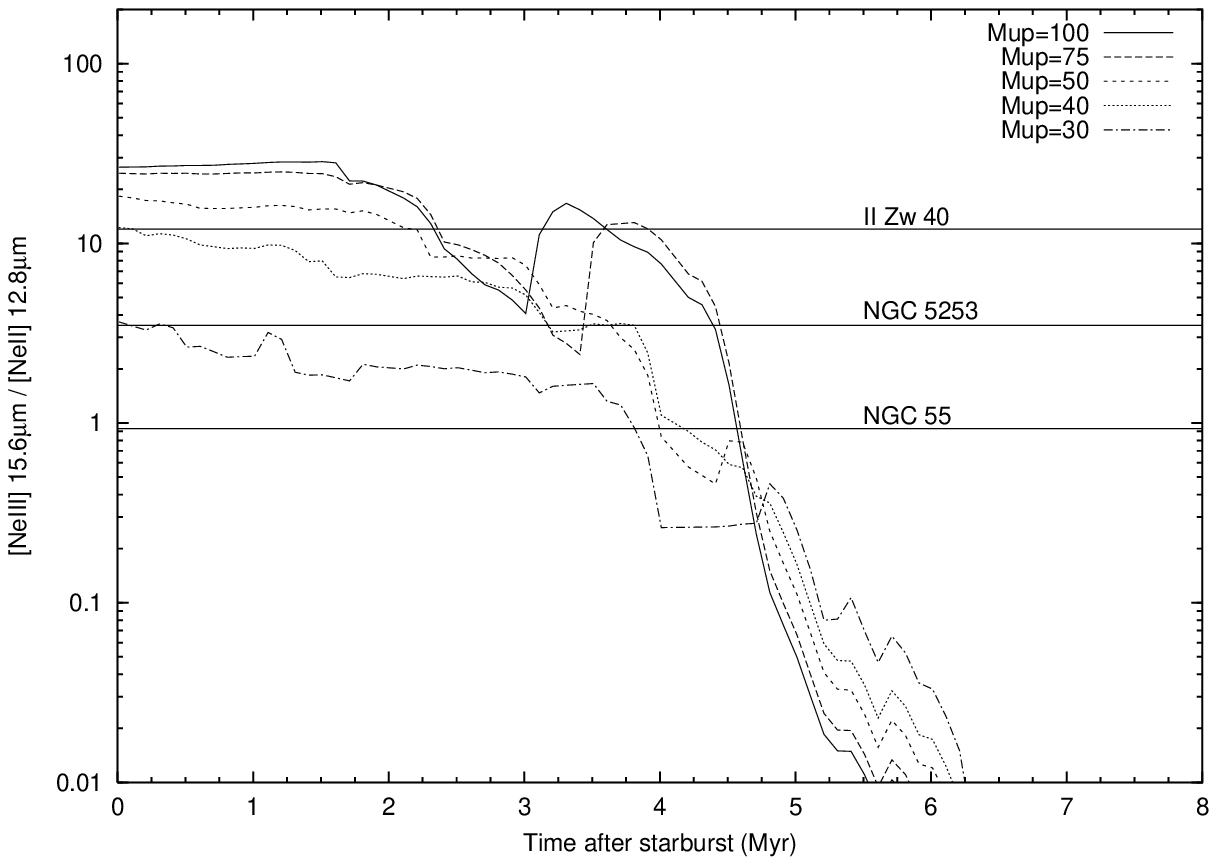,height=2in,angle=0,clip=t}
\epsfig{figure=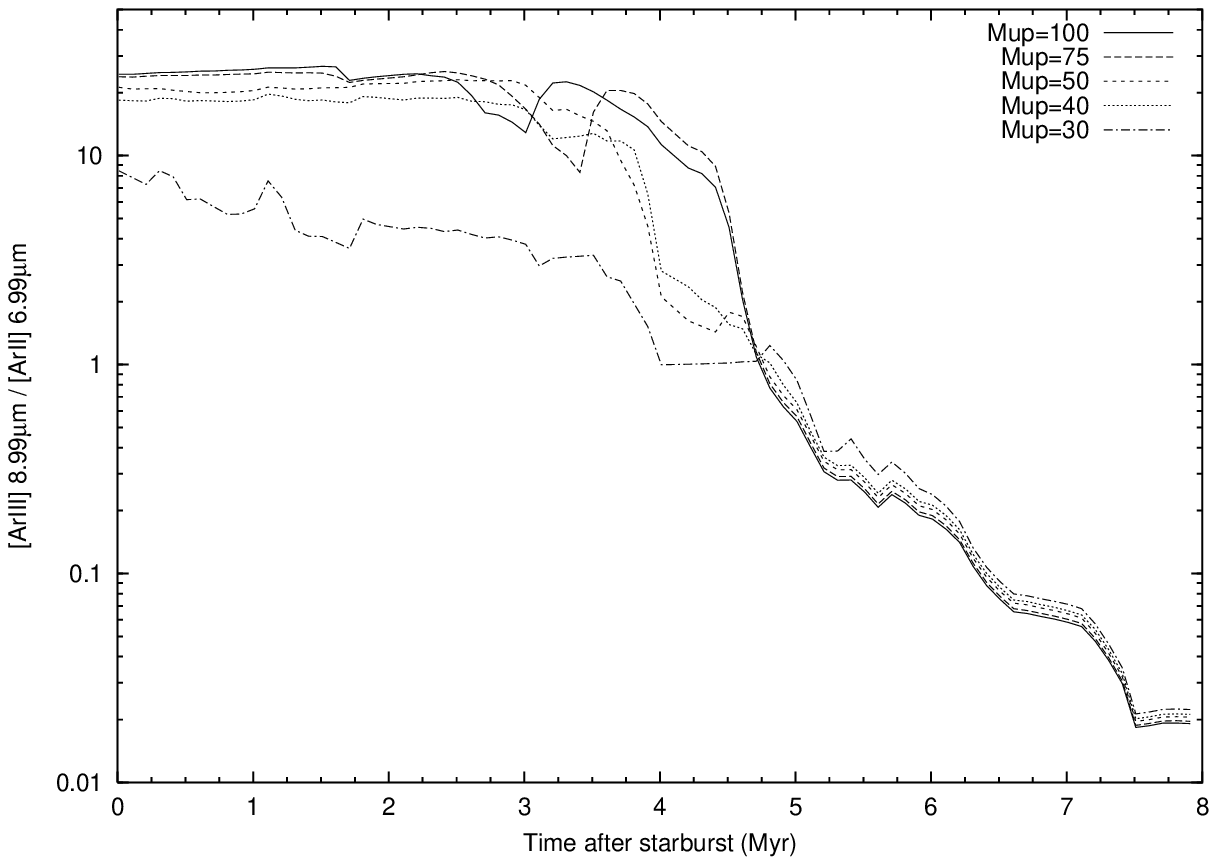,height=2in,angle=0,clip=t}
\epsfig{figure=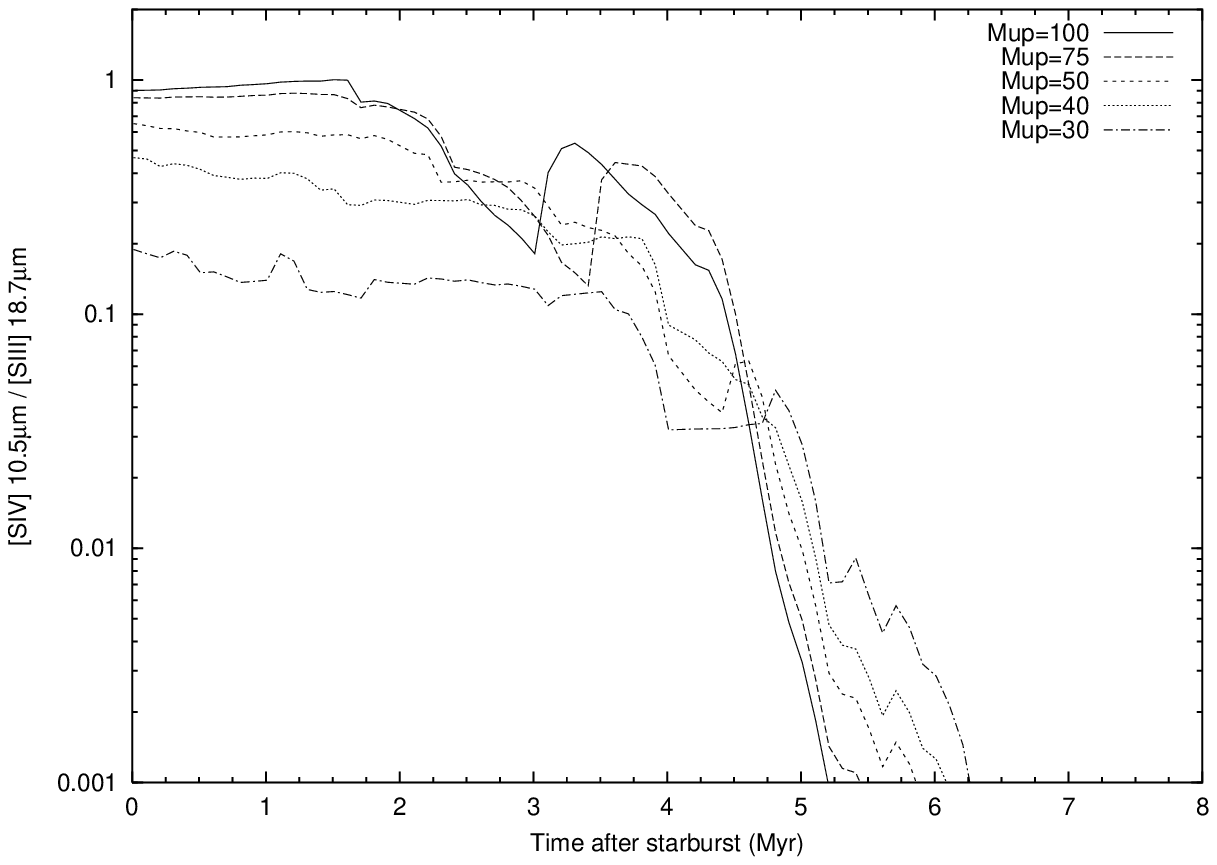,height=2in,angle=0,clip=t}
\hspace*{0.72in}
\epsfig{figure=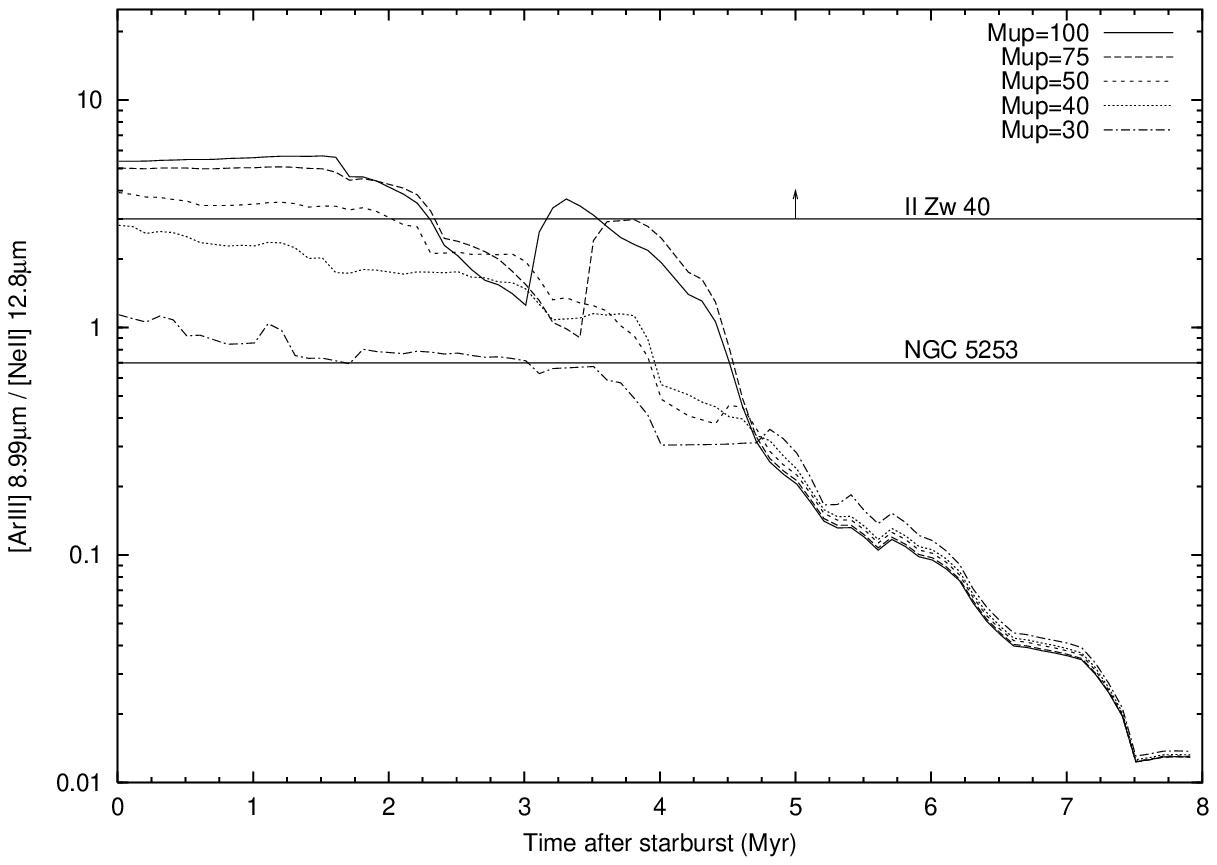,height=2in,angle=0,clip=t}
\epsfig{figure=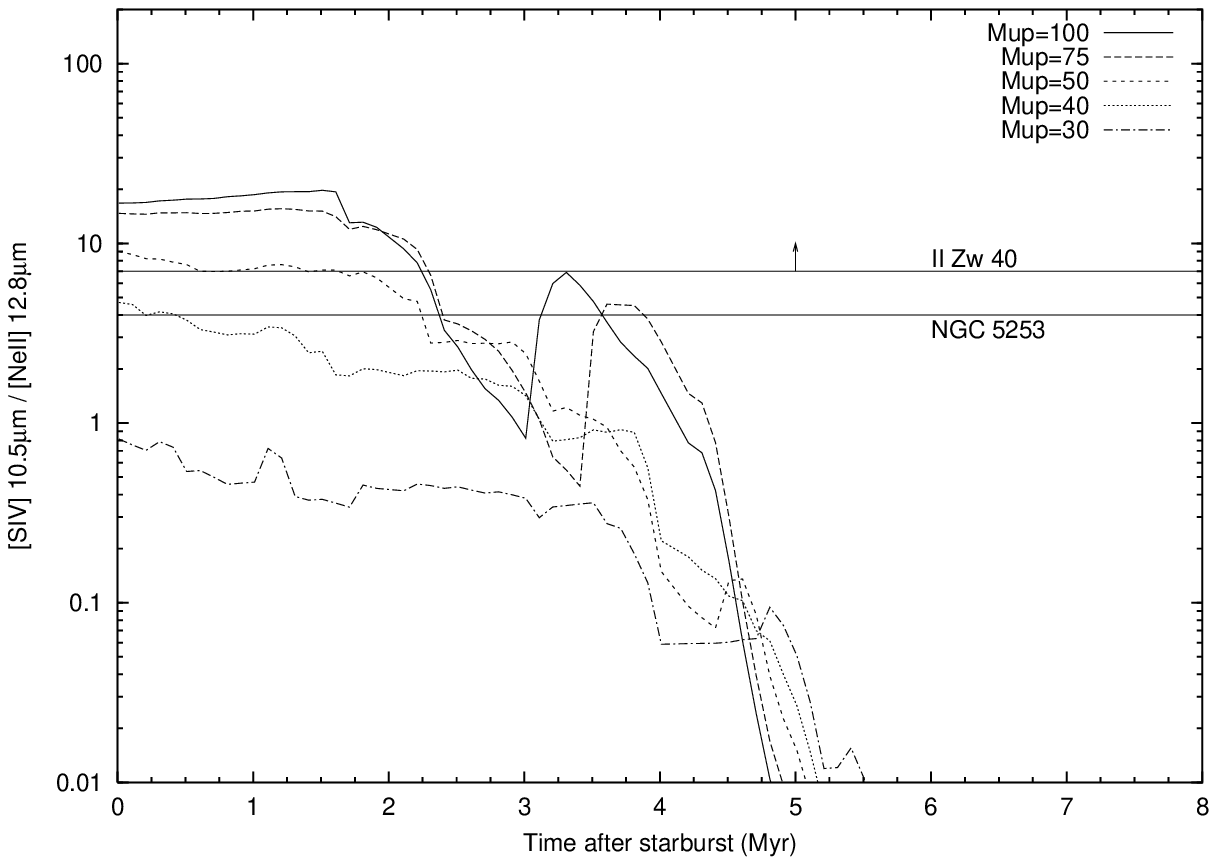,height=2in,angle=0,clip=t}
\figcaption{Line ratios versus time for low--metallicity Cloudy/Starburst99 models.  
Models assume
Z$=0.2$~Z$_{\sun}$, an instantaneous burst of star formation, and
an IMF with Salpeter slope and stars with masses between $1$~\Msol\ and \Mup.
Horizontal lines indicate observed line ratios for low--metallicity galaxies 
II~Zw~40, NGC~5253, and NGC~55, as given in table~\ref{tab-midir}.
}
\label{fig:lowZmodels}
\end{figure*}

Figure~\ref{fig:models} plots line ratios as a function of time for these simulations.
Table~\ref{tab-s99} summarizes the spectral synthesis models with \Mup$=100$~\Msol\
and compares to line ratios from the single--star models.  
Line ratios versus time for the low--metallicity models are shown 
in figure~\ref{fig:lowZmodels}.

Figure~\ref{fig:models} shows that in the 
first $2$~Myr, the mid--infrared line ratios fall from an initial plateau.
By $2$~Myr, the O3 through O5 dwarfs ($\Teff>44,500$~K) 
stars in the models have left the main sequence; by $2.5$~Myr, no O3 or O4 star of any 
luminosity class remains.  Wolf--Rayet stars, together with the remaining main sequence 
stars, create a second period of relatively high line ratios from 
$3.5$ to $5$~Myr.\footnote
{The Starburst99 model for $t=3.0$~Myr and M$_{up}=100~$\Msol\ predicts line ratios
that are sharply discontinuous from  ratios at $2.8$, $2.9$, $3.1$, and $3.2$~Myr.  
(The [\NeIII]/[\NeII] spike is 25 times higher than the surrounding points.)  The
M$_{up}<100$~\Msol\ models and sub-solar metallicity models have no spike.  
Though we have been unable to
pinpoint the cause from the Starburst99 output, we feel the $3$~Myr spike is spurious, 
not a physical effect, and we have removed it from the figures.}  
While the line ratios predicted for the Wolf--Rayet phase are lower than predicted for 
WR--only nebulae, clearly the Wolf--Rayet stars are important:  they produce
a renaissance of high line ratios after the O~stars have left the main sequence.
It does not make sense to parameterize a mid--infrared line ratio 
as though the flux came from a single main sequence star; the ensemble of stars, 
including the Wolf--Rayets, must be considered.

The models of \citet{thornley}, which otherwise used similar input spectra and 
nebular parameters to this work, did not include Wolf--Rayet stars.  As a result, 
[\NeIII]/[\NeII] drops monotonically with time in their figures~6 and 10, 
while our solar--metallicity curves (figure~\ref{fig:models}) are double--peaked. 

%

%
\section{DIAGNOSTICS OF STELLAR \Teff\ IN STARBURST GALAXIES}
\label{sec:diags}

\subsection{Approaches to Estimating \Teff}

In general, line ratios capable of indicating \Teff\ also depend
on metallicity, electron temperature, density, ionization parameter,
and the morphology of the ionized regions. Without constraints
on these other parameters, \Teff\ can be difficult to determine 
(e.g., \citet{morisset-apj}).  However, in the extreme conditions
in starbursts, we expect less range in ionization parameter and
morphology than in broad samples of \HII\ regions, and starburst
metallicities can be constrained by other line ratios. Therefore,
it is plausible that useful constraints on \Teff\ can be 
derived for these regions. We return to this topic in \S~\ref{sec:depends}.

At optical wavelengths, the ratio of [\OIII]~$5007$ to \Hbeta\ 
is frequently used as a diagnostic since both lines are easily observed
(see for example \citet{SL96}.)
They are relatively close in wavelength, and Balmer ratios can be 
used to correct for residual extinction.  However, large optical depths of 
interstellar extinction can make [\OIII]/\Hbeta\
reflect the conditions of the outer skin of starbursts only.  
Thus, [\OIII]/\Hbeta\ may not indicate the average conditions
throughout a highly--extincted starburst.  Also, a ratio 
composed of a forbidden metal line and a hydrogen recombination line 
is particularly sensitive to the metallicity, electron temperature, and 
density of the nebular region.
Additionally, [\OIII] can be shock excited \citep{cygloop}.

\citet{vp} have proposed an all--forbidden line diagnostic
$\eta^{\prime}$, which uses lines of [$\OII$], [$\OIII$], [$\SII$], and 
[$\SIII$] with wavelengths from $3726$~{\AA} to $9532$~{\AA}.  
This diagnostic was initially reported to work well for \HII\ 
regions \citep{kbfm}, but it is sensitive to morphology and 
shocks \citep{oey}.  Moreover, the diagnostic is poorly suited to 
starburst galaxies because it involves red lines that are seldom observed, 
and is extremely subject to reddening.

More robust optical line diagnostics can be made by comparing strengths
of helium recombination lines with 
recombination lines of hydrogen 
(Kennicutt \etal\ 2000; Ho, Filippenko, \& Sargent 1997; Doherty \etal\ 1995).
\nocite{kbfm,ho3,doherty95}
\Heopta~\AA\ and \Heoptc~\AA\ are attractive for this purpose because
their proximity to H$\alpha$ and H$\beta$, respectively, reduces 
reddening effects. However, these diagnostics still sample only 
the outer skin of the starburst, and the helium lines are weak, 
as discussed further in \S~\ref{sec:optical}.

Because they suffer less extinction, infrared spectral diagnostics
probe more deeply into a starburst than optical ones.
For example, $10$ magnitudes of extinction at $5500$~\AA\ corresponds to
only $1.1$ magnitudes at $2.2$~\micron, and $0.8$ magnitudes at $10$~\micron\
\citep{rl1985}.   The mid--infrared fine structure lines
are the most successful tools in this spectral region to estimate \Teff\ 
\citep{roche,kunze,al,thornley,fs-m82}.
These lines are less dependent on electron temperature than optical
forbidden lines.  However, their atomic constants are not well known
\citep{f97,f01,vanhoof,fe-proj-co}, and they are still sensitive to
metallicity and ionization parameter (see figure~10 of Thornley \etal\ 2000).
\nocite{thornley}

Because of the lack of strong atomic lines, attempts to use 
near--infrared lines to measure \Teff\ have 
focused on recombination lines of helium and hydrogen.
For stellar ionizing sources, the hardness
of the ionizing continuum determines the volume of He$^+$ relative
to H$^+$ \citep{osterbrock}.
For $\Teff > 40,000$~K, the He$^{+}$ and H$^{+}$ regions coincide;
the Str\"{o}mgren radii are 
approximately equal.  For lower \Teff, the zone of ionized H extends 
beyond the central zone of singly--ionized He (see 
figures~2.4 and 2.5 of Osterbrock (1989), and figure~1 of 
Shields (1993)).  Thus, by measuring the 
relative volumes of  He$^+$ and H$^+$ within a nebula, one can constrain 
the effective temperature of the ionizing stellar source(s).

The line ratio of $\HeIk / \Brgam$ has been used to 
estimate \Teff\ in starbursts (Doyon, Puxley, \& Joseph 1992; Doherty \etal\ 1995)
\nocite{dpj} \nocite{doherty95}
and planetary nebulae (Lumsden, Puxley, \& Hoare 2001a).
\nocite{lph320}
However, the strength of the $\HeIk$ $2^1P \rightarrow 2^1S$ line  
is not determined simply by recombination cascade, but also by
the population in the $2^1$P state \citep{shields}.  This level is pumped
from the ground state by $\lambda = 584$~{\AA} photons in the resonance 
transition $1^1S \rightarrow 2^1P$ \citep{shields,bs}.  
Photoionization of hydrogen, dust absorption, or Doppler shifting 
can change the resonance efficiency and thus the occupation of 
the $2^1P$ state.  The state can be further populated by collisions 
from the triplet states, primarily from $2^3S$ \citep{shields}.
A small $\HeIk / \Brgam$ ratio should indicate 
a soft continuum where there are few $584$~\AA\ photons and few 
helium recombinations. Otherwise, the ratio is likely to be a poor 
measure of starburst \Teff\ due to the dependence on nebular dust content, 
electron temperature, and density, as well as on the ionizing continuum. 
Some of this complex behavior is seen in photoionization models 
(figure~1d of Shields 1993). \nocite{shields}

The $\HeIh / \Brten$ ratio was proposed as a
\Teff\ diagnostic by \citet{vanzi}, and has been measured in several 
starburst galaxies 
(Vanzi \etal\ 1996; Vanzi \& Rieke 1997; Engelbracht, Rieke, \& Rieke 1998;
 F\"{o}rster Schreiber \etal\ 2001)  
\nocite{vanzi} \nocite{vr}  \nocite{chad}  \nocite{fs-m82}
and planetary nebulae  (Lumsden, Puxley, \& Hoare 2001b).
\nocite{lph328}
The \HeIh\ line and \Brten\ are close in wavelength, 
and A$_{1.7~\micron}$ is only one-sixth of A$_V$, making their 
ratio nearly reddening-independent and also allowing the photons to escape 
from relatively obscured regions. 
Unlike \HeIk, the \HeIh\ $4^3D \rightarrow 3^3P^0$ transition arises 
almost entirely from recombination cascade.  The relevant levels are 
triplet states, so they cannot be pumped from the ground state, because an 
electron spin flip would be required \citep{bs}.  As a result, the 
line ratio is insensitive to nebular conditions, and is determined almost 
entirely by the relative sizes of the H and He ionization zones.

Figure~8 of \citet{fs-m82} plots the behavior of the \HeIh/\Brten\ 
ratio as a function of \Teff, as predicted by photoionization models 
for a starburst environment ionized by hot main sequence stars.  
The \HeIh/\Brten\ ratio is small 
for  \Teff$ < 30,000$~K because there are many more photons capable of
ionizing hydrogen (ionization potential of $13.6$~eV) than neutral 
helium (ionization potential of $24.6$~eV).  For \Teff $>30,000$~K,
the ratio rapidly increases as the zone of singly--ionized helium 
overlaps more of the hydrogen Str\"{o}mgren sphere.  The ratio then
saturates for \Teff $> 40,000$~K, as the He$^+$ and H$^+$ regions coincide.
For $n_e = 100$~{\cc}, the saturated ratio is
\begin{equation}
\HeI\ 1.7 / \Brten\ = 3.60~C_{1.7}~[n(He) / n(H)] ,
\label{eq:saturated}
\end{equation}
where $n(He)/n(H)$ is the gas--phase abundance of helium (by number) 
relative to hydrogen, and the term C$_{1.7}$ expresses the weak
dependence on electron temperature \citep{vanzi}.  
For  T$_e = 10^4$~K, C$_{1.7} = 1.000$; other values are 
listed in table~\ref{tab:c17}.
The helium abundance $n(He)/n(H)$ increases from the primordial 
value of approximately  $0.08$ \citep{ih-he,bono}
to $0.1$ for the Milky Way.  
Thus, \HeIh/\Brten\ should saturate at a value of $0.27$ to $0.38$.

The \Teff\ diagnostics discussed above do not necessarily agree.
For example, in the starburst galaxy  He~2--10,  [\OIII]/\Hbeta\
and [\OI]/\Halpha\  indicate \Teff$>39,000$~K \citep{he2-10opt}, whereas 
mid--infrared line ratios indicate \Teff$<37,000$~K \citep{roche}.  The
\HeIk/\Brgam\ observed by \citet{vr} would indicate \Teff$=39,000$~K using
the conversion of \citet{dpj}.  At poor signal--to--noise, \citet{vr}
measure \HeIh/\Brten\ and find \Teff$=36,000$~K.  
This few thousand Kelvin disagreement translates into 
a serious disagreement in stellar mass:
a \Teff\ of $36,000$~K corresponds to approximately an O8V spectral type,
which from eclipsing binaries should have a mass of $\sim 22$ to $25$~\Msol\ 
\citep{andersen,ostrov,niemela,gies}; whereas 
a \Teff\ of $40,000$~K corresponds to an O6.5V to O7V spectral type, 
which should have a mass of $\sim 35$~\Msol\ \citep{gies,niemela}.

We now test \Teff\ diagnostics against each other in light of
the stellar synthesis models detailed above.  Because the line
physics of \HeIh/\Brten\ is simple and well--understood (see 
\S~\ref{sec:diags}), we assume this diagnostic is unbiased, and thus
accurately reflects the ionizing continuum,
within the limitations of measurement error.

\subsection{Testing \HeIk/\Brgam}
\label{sec:heh_hek}
In this section we consider the galaxies for which we obtained 
\HeIh/\Brten\ measurements, as well as three galaxies with
\HeIh/\Brten\ measurements  available in the literature: 
NGC~253 \citep{chad}, for which the stellar continuum 
was subtracted as in this work; M82 \citep{fs-m82}, for which 
representative stellar spectra were subtracted; and NGC~5253 \citep{vr}, 
for which the stellar continuum is weak enough to ignore.  These 
three galaxies, together with the six galaxies for which we observed
\HeIh/\Brten, we term our expanded sample.  We also take measurements
of \HeIk/\Brgam\ from the literature for the galaxies
in the expanded sample.

\begin{figure*}
\figurenum{5}
\plotone{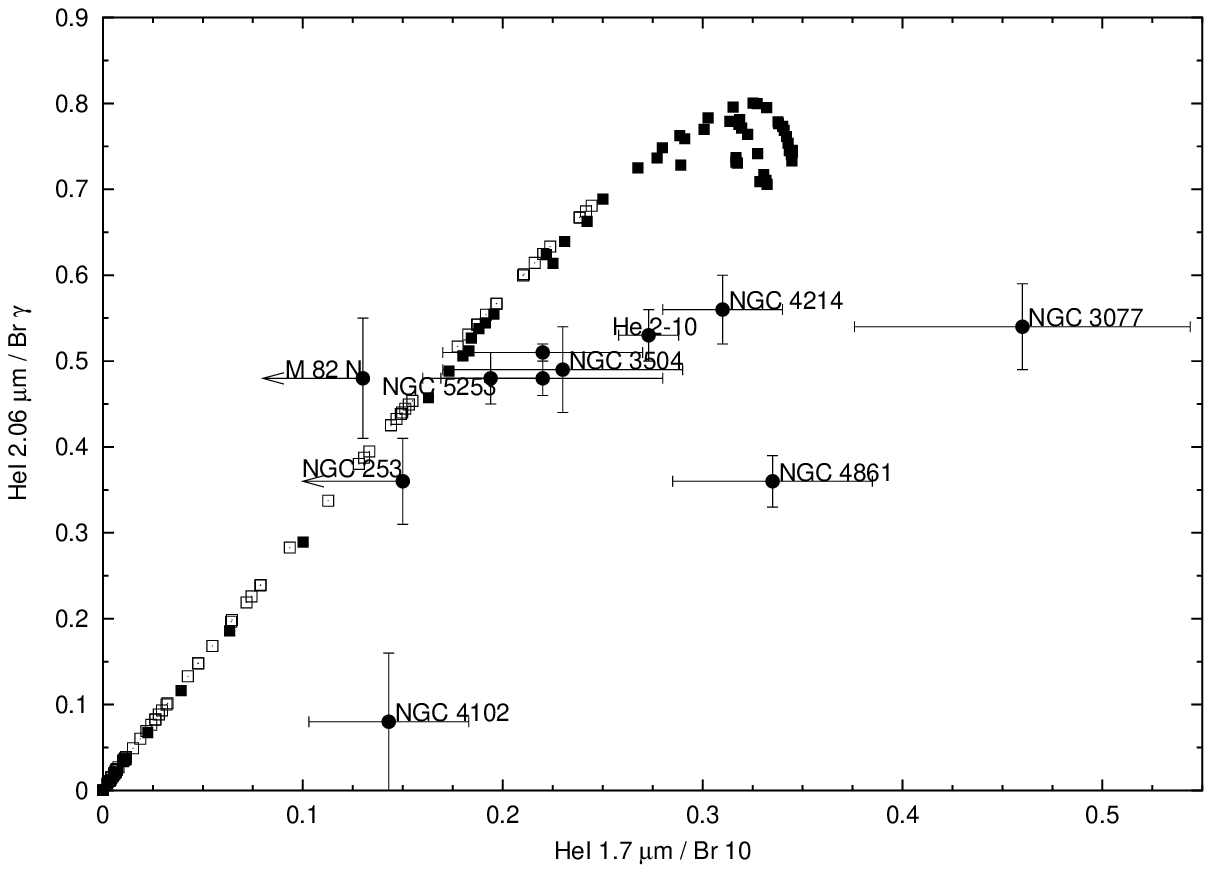}
\figcaption{Comparison of the near--infrared line ratios.  Included are
the galaxies we observed
as well as NGC~5253 \citep{vr,lpd}, NGC~253 \citep{chad}, and three regions 
of M82 \citep{fs-m82}.  The M82 regions are:  the nucleus, 
labeled ``M82~N''; the \Brgam\ source ``B1'' located 
$10$\arcsec\ southwest of the nucleus (the upper unlabeled point in this plot); 
the \Brgam\ source ``B2'' located $5$\arcsec\ southwest of the nucleus
(the lower unlabeled point.)  Coordinates for the M82 regions are given
in the footnotes to table~\ref{tab-midir}.
Filled squares show line ratio values from Starburst99/Cloudy models with
solar metallicity and \Mup$=100$~\Msol, run every $0.1$~Myr after an 
instantaneous burst.  The open squares represent models with \Mup$=30$~\Msol.
Lowering the metallicity changes the tracks insignificantly.
}
\label{fig:heh_hek}
\end{figure*}

Figure~\ref{fig:heh_hek} plots \HeIk/\Brgam\
versus \HeIh/\Brten.  NGC~3077, NGC~4861, NGC~4214, and He~2--10 all have 
\HeIh/\Brten\ ratios consistent with the saturated value of $\approx 0.3$, 
within the measurement errors and the expected variation of helium abundance.  
Thus, these starburst regions appear to contain massive stars 
(\Teff $> 39,000$~K if main sequence stars.) 
By contrast, NGC~253, NGC~4102, and the nucleus of 
M82 have \HeIh/\Brten$<0.15$, and thus are inferred to have softer
ionizing continua (\Teff $\lesssim 37,000$~K if main sequence stars.)
NGC~3504 and the two off--nuclear regions of M82 have line ratios
intermediate to these extremes.

Figure~\ref{fig:heh_hek} illustrates that \HeIk/\Brgam\ does not trace
\HeIh/\Brten\ as the models predict.
The nucleus of M82 demonstrates that \HeIk\ may be strong while \HeIh\ is weak,
contrary to the expected behavior (but expected if \HeIk\ is  pumped.) 
However, for most galaxies, \HeIk\ is \emph{too weak} for the measured \HeIh.  
This is the first direct demonstration that \HeIk/\Brgam\ is a poor 
diagnostic of \Teff\ in 
starburst galaxies.  Radiative transfer considerations have   
predicted that the behavior of \HeIk\ should not be a simple function
of \Teff\ \citep{shields}.  \citet{lph320} confirm this complex behavior 
for planetary nebulae, though they attempt to constrain the dependence
on T$_e$ and density by also considering optical \HeI\ lines \citep{doherty95}.
However, the data do not contradict the expectation that a very low 
\HeIk/\Brgam\ ratio (below $\sim0.2$) 
indicates that the continuum is fairly soft, because there
would be few ionizing photons and also few resonantly scattered photons.

\begin{figure*}
\figurenum{6}
\plotone{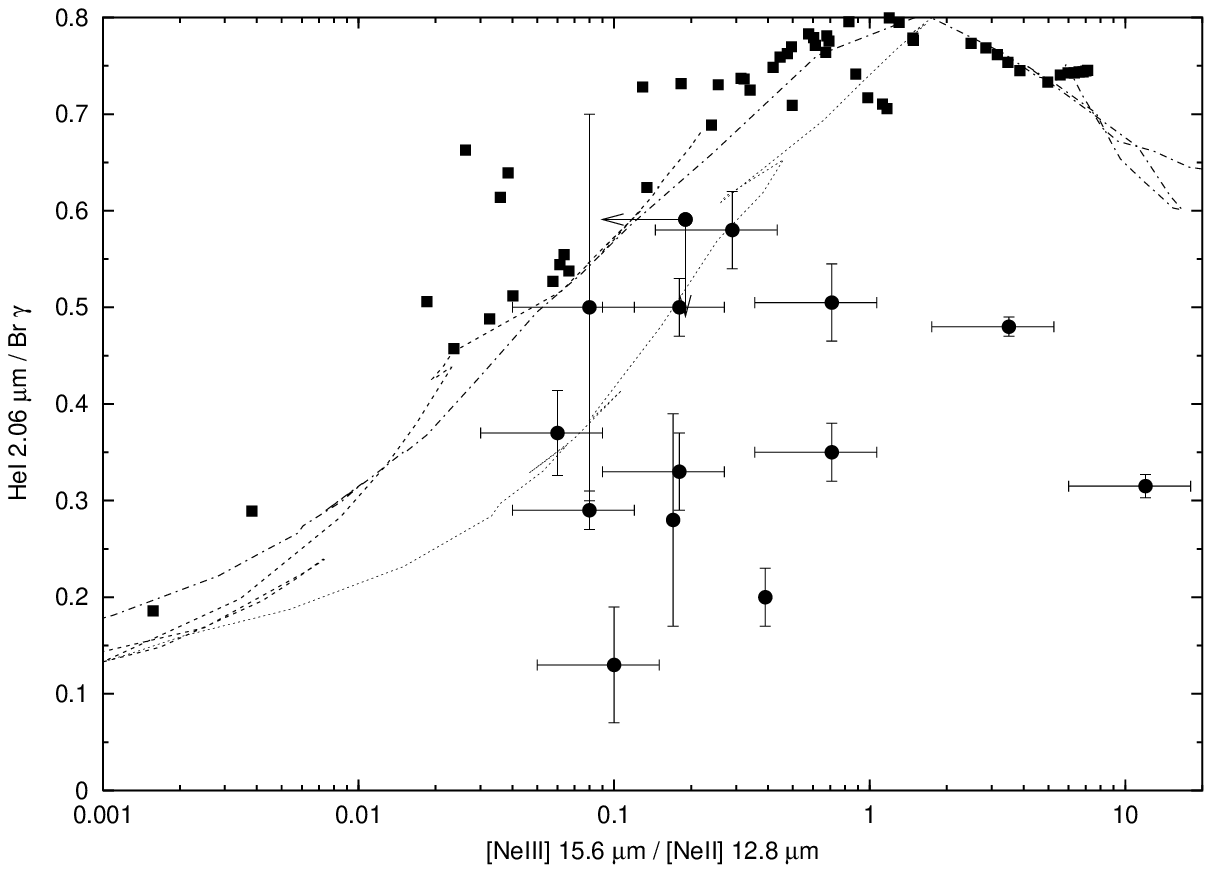}
\figcaption{Comparison of [\NeIII]/[\NeII] with \HeIk/\Brgam.
Mid--infrared ratios are from \citet{thornley}, and \HeIk/\Brgam\
measurements from the 
literature \citep{doherty95,dpj,chad_thesis,fs-m82,genzel,lester,lpd,schinnerer,vanzi}.
Starburst99/Cloudy models are overplotted:  filled squares show 
solar--metallicity models with \Mup$=100$~\Msol;
the dashed line shows solar--metallicity models with \Mup$=30$~\Msol;
the dot--dashed line shows the low--metallicity, \Mup$=100$~\Msol\ models;
and the finely dotted line shows the low--metallicity, \Mup$=30$~\Msol\ models.
}
\label{fig:ne_206}
\end{figure*}

We further consider the reliability of the \HeIk/\Brgam\ ratio in 
figure~\ref{fig:ne_206}, by comparing it to the mid--infrared 
line ratio [\NeIII]~15.6\micron/[\NeII]~12.8\micron.  Here, too,
\HeIk/\Brgam\ is too low for a given [\NeIII]/[\NeII] (compared to model
predictions) and there is no obvious correlation between the two ratios.
An alternative interpretation of figure~\ref{fig:ne_206} would be that
\HeIk/\Brgam\ is correct and [\NeIII]/[\NeII] is systematically 
overproduced;  we feel this is unlikely because, as we will 
demonstrate in \S~\ref{sec:midir-test}, [\NeIII]/[\NeII] is 
\emph{underproduced} in starburst galaxies with respect to 
the predictions of a Salpeter IMF extending to $100$~\Msol.

%
%
\subsection{Testing Optical \Teff\ Indicators}
\label{sec:optical}
How well do optical forbidden and recombination line ratios estimate 
\Teff\ in starbursts?  Figures~9 and 10 of \citet{kbfm} show
that the recombination ratios \Heoptb/\Hbeta\ 
and \Heopta/\Halpha, as well as [\OIII]/\Hbeta, all track
\Teff\ well in Milky Way, LMC, and SMC \HII\ regions, where \Teff\
could be determined by classifying all the ionizing stars.
How well do these diagnostics perform in starburst galaxies?

\begin{figure*}
\figurenum{7}
\plotone{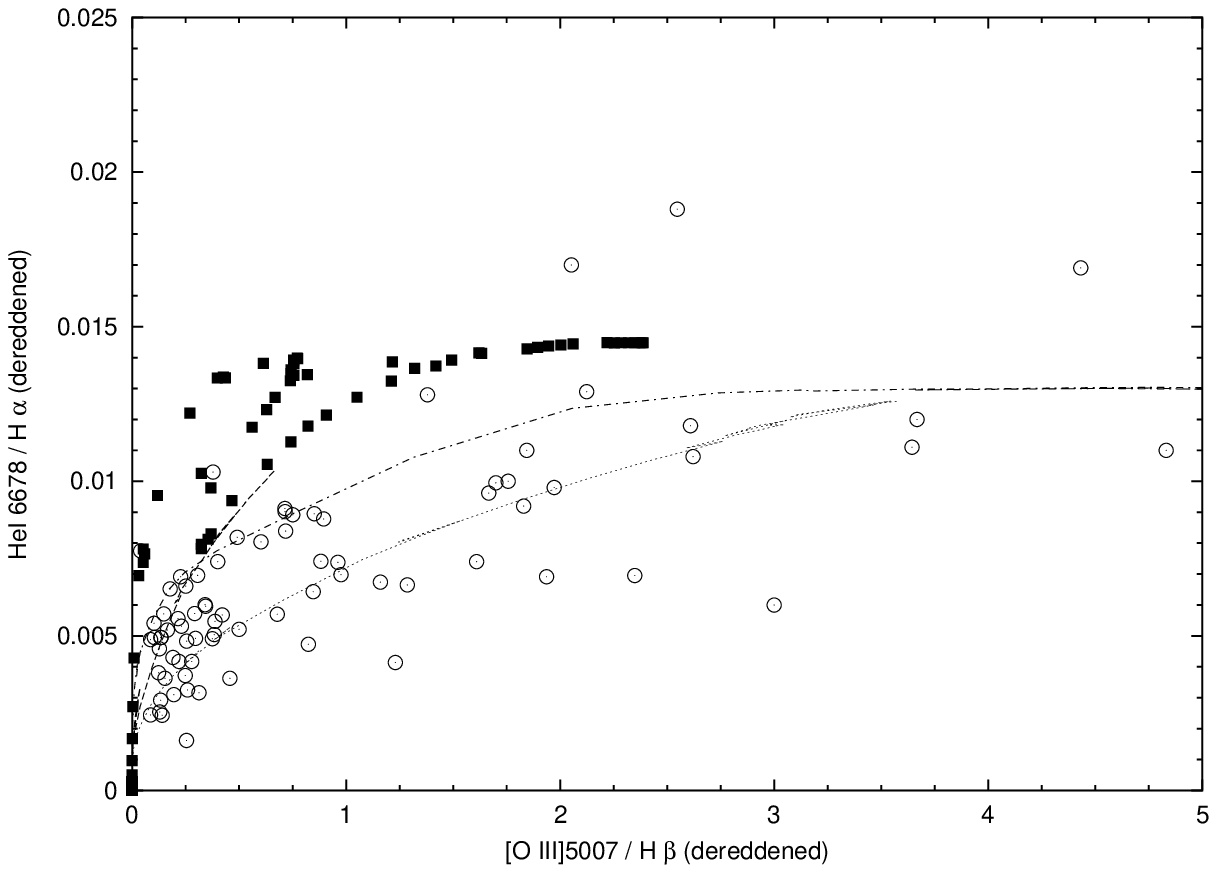} 
\figcaption
{Testing optical indicators of \Teff.
Circles represent the subsample of galaxy nuclei that 
\citet{ho3} classify as being ``\HII\ region--like''. 
Line fluxes are from \citet{ho3}, dereddened by us following the 
extinction law of Cardelli, Clayton, \& Mathis (1989), 
with $R_V = 3.1$.
Starburst99/Cloudy models are represented as in figure~\ref{fig:ne_206}.
Note that the $[\OIII]/\Hbeta$ ratios tabulated in \cite{ho3}
are incorrectly labeled as dereddened, when in fact they are
observed values.
}
\label{fig:optical}
\end{figure*}

In figure~\ref{fig:optical}, using dereddened data from \citet{ho3}, 
we compare the behaviors of \Heopta/\Halpha\ and [\OIII]/\Hbeta\ 
in nuclear starbursts to the predictions of Starburst99/Cloudy 
photoionization models.  
Galaxies with [\OIII]/\Hbeta$<0.5$ generally have low \Heopta/\Halpha, 
indicating general agreement that \Teff\ is low in these galaxies.
At higher line ratios, there is considerable scatter.
For most of the plotted galaxies, [\OIII]/\Hbeta\ is systematically high for 
a given \Heopta/\Halpha, compared to a solar--metallicity, 
\Mup$=100$~\Msol\ track.  Lowering the metallicity of the model reduces but 
does not eliminate the disagreement between diagnostics.  
Only an extreme model (low metallicity,
\Mup$=30$~\Msol) can fit the data reasonably well.  

A possible explanation would be that [\OIII] in starbursts is often shock--excited 
by supernovae \citep{cygloop}, which would be a rare effect in \HII\ regions and 
thus not affect the \citet{kbfm} plots. 
In particular, [\OIII]/\Hbeta\ values above $\sim 2.5$ require 
sub--solar metallicity or  excitation by shocks.  Thus, 
figure~\ref{fig:optical} suggests that [\OIII]/\Hbeta\
is systematically high or \Heopta/\Halpha\ is systematically low in
starburst galaxies.

\begin{figure*}
\figurenum{8}
\plotone{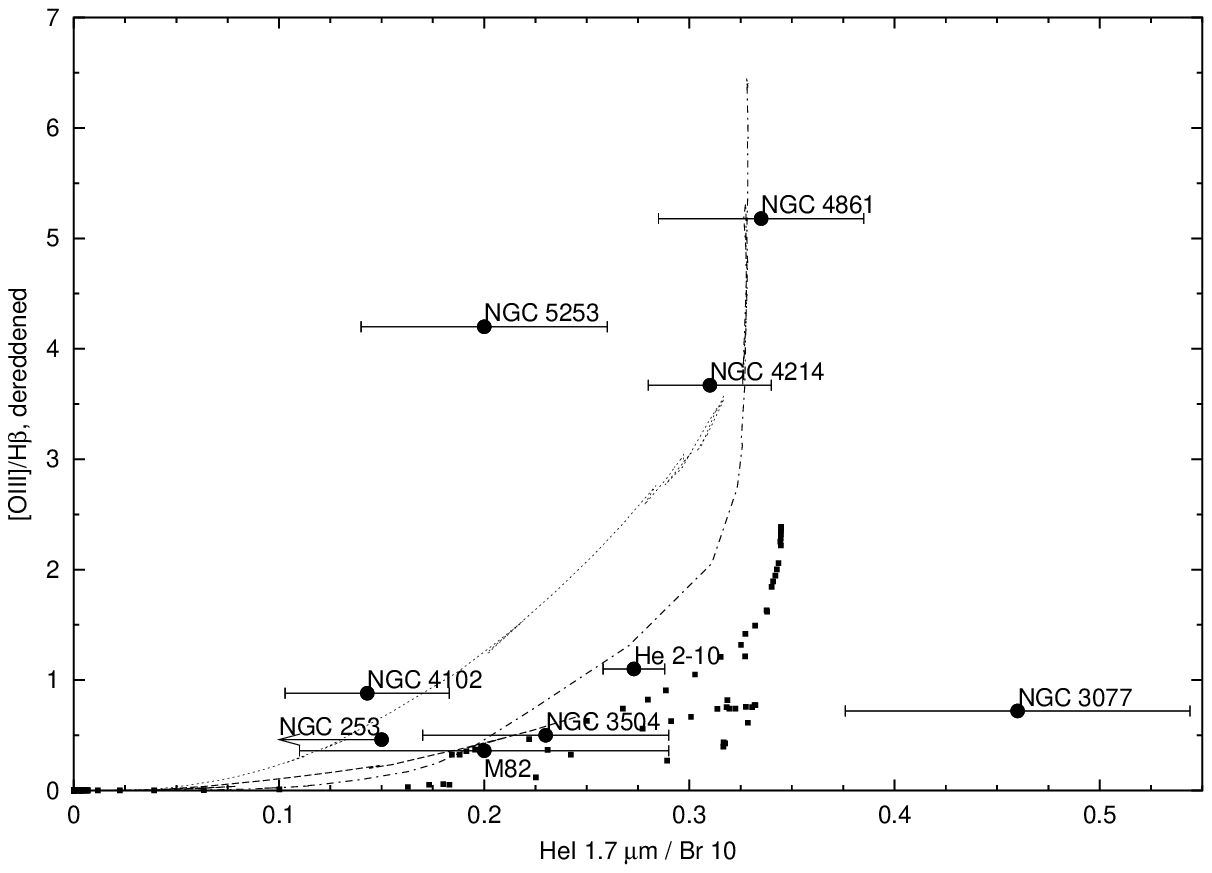}
\figcaption{Testing $[\OIII]/\Hbeta$ against $\HeIh/\Brten$. 
The filled circles are measurements from the literature:
dereddened $[\OIII]/\Hbeta$ measurements
were taken from \citet{op,vvgds,vc}; we also took observed
$[\OIII]/\Hbeta$ ratios from \citet{ho3} and dereddened
them following \citet{ccm}.  For M82, we use the dereddened value of
\citet{ho3}, averaged with observed values from Armus, Heckman, \& Miley 
(1989) and \citet{wzxd}.  Starburst99/Cloudy models are represented
as in figure~\ref{fig:ne_206}.
}
\label{fig:o3_heh}
\end{figure*}

As a further test, figure~\ref{fig:o3_heh} 
plots [\OIII]/\Hbeta\ versus \HeIh/\Brten\ for our expanded sample.
Overplotted are Starburst99/Cloudy models as in figure~\ref{fig:optical}.
For low values of [\OIII]/\Hbeta, the error bars are too large to judge
whether the two diagnostics correlate.
As in figure~\ref{fig:optical}, the highest [\OIII]/\Hbeta\
values observed require sub--solar metallicity or shock excitation of [\OIII].

\begin{figure*}
\figurenum{9}
\plotone{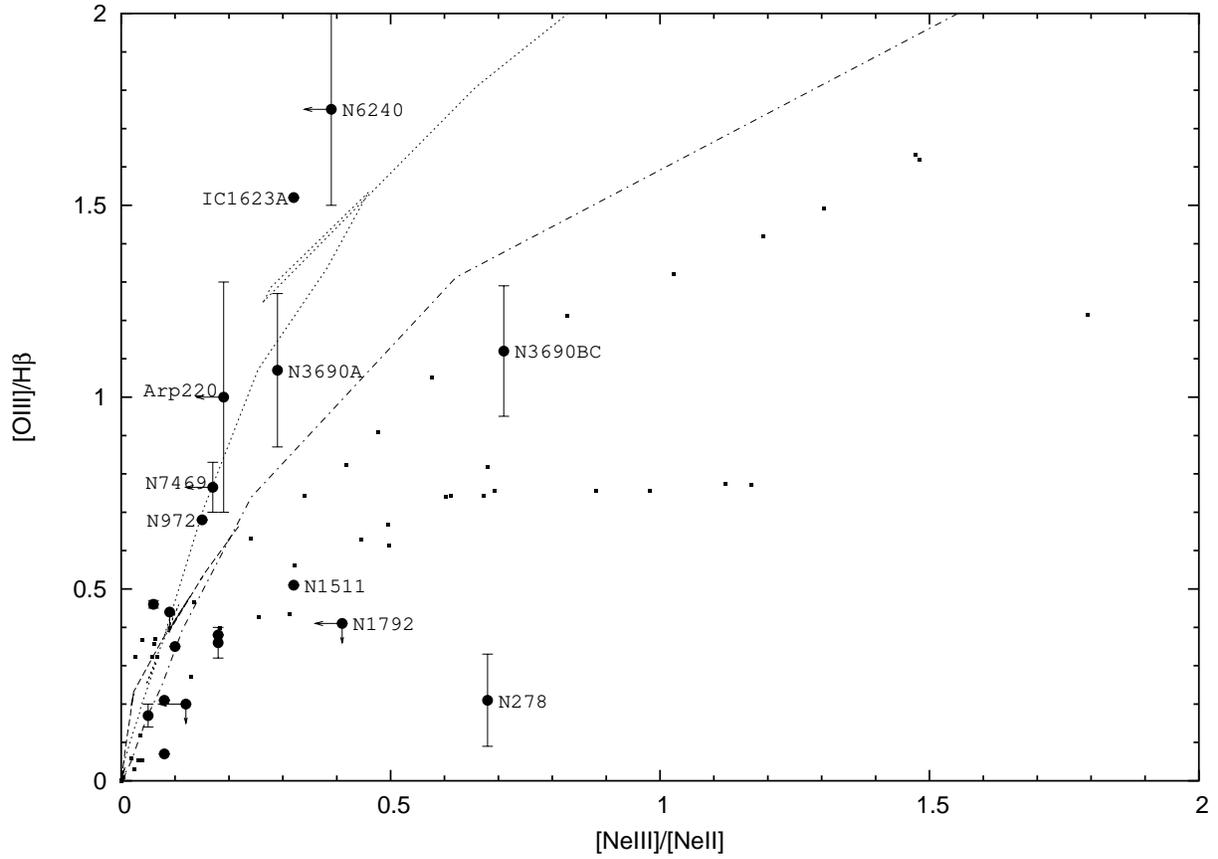}
\figcaption{
[\OIII]/\Hbeta\ from the literature versus [\NeIII]/[\NeII]
from \citet{thornley} (filled circles).  
Starburst99/Cloudy models are represented as in figure~\ref{fig:ne_206}.
For clarity, we omit labels for the galaxies with smallest line ratios.
In order of increasing [\OIII]/\Hbeta\ flux (or flux upper limit), they are:
IC~342, M83, NGC~986, NGC~7552, NGC~6946, M82, NGC~3256, NGC~4945,
and NGC~253.
}
\label{fig:o3_Ne}
\end{figure*}

Next, we examine the behavior of [\OIII]/\Hbeta\ versus [\NeIII]/[\NeII]
in figure~\ref{fig:o3_Ne} (omitting for now NGC~5253, II~Zw~40, and NGC~55 
because of their low metallicity.)  Optical line ratios are from the literature, 
and \emph{ISO} observations of [\NeIII]/[\NeII] are from \citet{thornley}.
Galaxies with [\OIII]/\Hbeta$<0.5$ generally have line ratios consistent with the
overplotted Starburst99/Cloudy models.  With higher [\OIII]/\Hbeta,
the scatter increases.  Without accurately knowing the metallicity of each
galaxy in figure~\ref{fig:o3_Ne}, it is difficult to judge how much of
the scatter in [\OIII]/\Hbeta\ versus [\NeIII]/[\NeII] is due to the
sensitivity of [\OIII]/\Hbeta\ to metallicity rather than effective temperature.

According to the Cloudy models, metallicity alone cannot explain the line ratios
of NGC~6240, IC~1623A, Arp~220, NGC~3690A, and NGC~7469 (and possibly NGC~972) 
in figure~\ref{fig:o3_Ne}.  Low metallicity and a upper mass cutoff of 
$30$~\Msol\ could together explain all but IC~1623A and NGC~6240.  Alternatively,
aperture mismatch, severe extinction, or shock excitation of [\OIII] could be at 
work.  NGC~972 is not  
strongly centrally concentrated in optical emission line images, so 
the explanation may lie in aperture mismatch: the optical line ratios were measured 
with slitwidths of a few arcseconds, while the \emph{ISO} neon lines were measured with 
a $14$\arcsec\ by $27$\arcsec\ aperture.  The remaining discrepant galaxies all
have very heavily obscured star formation regions,
and it is likely that the discrepancy arises because the optical and 
mid--infrared spectra sample distinctly different regions along the line of sight.
We also note that NGC~278 has extremely low [\OIII]/\Hbeta\ for its measured
[\NeIII]/[\NeII].  Higher--spatial resolution mid--infrared spectroscopy 
(e.g., with SIRTF) may resolve this discrepancy. 
We will delay discussion of whether [\NeIII]/[\NeII] is a 
reliable \Teff\ diagnostic until \S~\ref{sec:midir-test}.
 
Next, we consider the optical helium and hydrogen recombination
lines, which should form more accurate starburst \Teff\ diagnostics 
than a forbidden/recombination pair like [\OIII]/\Hbeta.  To reduce
reddening effects, we select \HeI\ lines close in wavelength to H lines.
Unfortunately, the helium lines are weak:  \Heopta\ saturates at 
$0.014$ of the strength of \Halpha, and \Heoptc\ saturates at $0.05$ of \Hbeta.
As such, in the spectral atlas of \citet{ho3}, \Heopta\ was 
detected in only $108$ of $418$ galactic nuclei, and \Heoptc\ 
in only $16$ nuclei.  The small sample indicates that \Heoptc\ is only 
marginally detected, and we do not consider it further.
 
Figures~\ref{fig:optical} and \ref{fig:o3_Ne} have already implicated
[\OIII]/\Hbeta\ as an unreliable \Teff\ indicator for 
[\OIII]/\Hbeta$\ga0.5$.  This makes it hard to gauge the reliability
of \Heopta/\Halpha\ in figure~\ref{fig:optical}.
Also, the sample sizes are too small to compare the optical recombination 
line ratios to [\NeIII]/[\NeII], \HeIh/\Brten, or \HeIk/\Brgam\
individually.  Instead, we use the latter three \Teff\ indicators
together to test how well the optical recombination line ratios
correlate with \Teff.  In table~\ref{tab-opt}, we list galaxies 
with measurements of at least two different \Teff\ indicators, 
in order of increasing \Teff, as determined from [\NeIII]/[\NeII], 
\HeIh/\Brten, and \HeIk/\Brgam\ (when $\le 0.2$), as available.  
Due to measurement error and uncertainty in the relative calibrations
of the diagnostics, the ordering is somewhat uncertain.  
The published plots of the \citet{ho3} spectra lack the dynamic range to 
assign upper limits to the undetected optical recombination lines.
These are marked as ``non det'' in table~\ref{tab-opt}.

In general, table~\ref{tab-opt} shows some correlation between 
\Heopta/\Halpha\ and \Teff, though with considerable scatter.
Using Kendall's~$\tau$
rank correlation test on the eight galaxies with measured \Heopta/\Halpha,
there is only a $5\%$ chance that \Teff\ and \Heopta/\Halpha\ are uncorrelated.

\subsection{Testing the Mid--Infrared Fine Structure Line Ratios}
\label{sec:midir-test}
In the mid--infrared, ratios of the fine structure lines 
[\NeIII]~$15.6$~\micron, [\NeII]~$12.8$~\micron, [\ArIII]~$8.99$~\micron, 
[\ArII]~$6.99$~\micron, [\SIV]~$10.5$~\micron, and [\SIII]~$18.7$~\micron\ 
have been used to test for the presence of hot stars in starbursts.  
From space, \emph{ISO} measured these lines at low spatial resolution 
($14$\arcsec\ by $27$\arcsec\ aperture for [\NeIII]/[\NeII]) 
\citep{thornley,fs-m82,kunze}. 
Ground--based observations \citep{roche,al}
provide higher spatial resolution, but only the [\SIV], [\ArIII], and [\NeII] 
transitions can be observed through the atmosphere.
As a result, ground--based studies must use heteronuclear line ratios, 
which are less ideal than homonuclear ratios available from space because
they are much more sensitive to elemental abundances and dust depletion.

In table~\ref{tab-midir}, we collect measurements of the mid--infrared line ratios
and \HeIh/\Brten\ in starburst galaxies, including five regions within M82.
M82 provides a testing ground for the accuracy of the mid--infrared line ratios 
as \Teff\ diagnostics; \emph{ISO} measured homonuclear line ratios in the 
center \citep{fs-m82}, and this region has been 
mapped at 1\arcsec\ resolution in [\NeII], [\ArIII], and [\SIV] \citep{al}, 
identifying the nucleus and three infrared--bright regions nearby (all regions
defined in the footnotes to table~\ref{tab-midir}.)

Based on the heteronuclear mid--infrared line ratios and \HeIh/\Brten, we  find 
that region W2 and the nucleus of M82 both require \Mup$<65$~\Msol, region E1
 requires \Mup$<60$~\Msol, and region W1 requires \Mup$<50$~\Msol.  
The heteronuclear and homonuclear mid--infrared line ratios and \HeIh/\Brten\ 
within the SWS/ISO aperture require \Mup$<50$~\Msol.  Models with \Mup$=100$, $75$, 
$70$, or $65$~\Msol\ do not produce the observed ratios in any of these regions.
Thus, we find that the heteronuclear line ratios give consistent ages and upper mass 
cutoffs  for individual regions near the center of M82, in agreement with \HeIh/\Brten,
and when averaged over the \emph{SWS/ISO} aperture, give answers consistent with 
the homonuclear line ratios.

We further test the mid--infrared line ratios using the five other solar--metallicity 
galaxies listed in table~\ref{tab-midir}.  For NGC~4102 and NGC~6240, the 
constraints are poor, and the line ratios can be fit by \Mup$=40$ to $100$~\Msol.  
In NGC~6946, \HeIk/\Brgam\  and [\NeIII]/[\NeII] disagree unless \Mup$<65$~\Msol,
but as we cautioned in \S~\ref{sec:heh_hek}, \HeIk\ is not a reliable diagnostic.
For NGC~253, \HeIh/\Brten\ and the neon ratio cannot be simultaneously matched by 
the \Mup$=100$~\Msol\ model, but models with \Mup$\le 75$~\Msol\ can fit the ratios.
For He~2--10, the line ratios require \Mup$<65$~\Msol, mostly because of low observed 
[\ArIII]/[\NeII].

Our conclusion is that in individual regions and entire starbursts, the different
heteronuclear and homonuclear mid--infrared line ratios and \HeIh/\Brten\ 
give consistent answers as to age and \Mup.  This agreement supports use of 
the mid--infrared line ratios as diagnostics of the ionizing radiation field.
The mid--infrared lines have large equivalent widths and a range of excitation 
energies, making them potentially powerful diagnostics.  

\section{DIAGNOSING IONIZING CONDITIONS IN STARBURSTS}
\label{sec:neon}

\subsection{Mid--Infrared Line Ratio Dependencies}
\label{sec:depends}

The mid--infrared line ratios depend on several physical parameters: metallicity,
ionization parameter, morphology, and the strength and shape of the 
ionizing continuum. To be confident in applying these ratios, we need to
disentangle these various effects. We consider each parameter in turn.

{\bf Metallicity}. As metallicity decreases, the relative high--excitation 
line emission increases, because lower--metallicity stars have harder spectra
and because lower--metallicity nebulae cool less efficiently.  
Another effect is that Wolf-Rayet stars require larger progenitor masses with 
decreasing metallicity.  These effects can be seen by comparing the 
low--metallicity models (figure~\ref{fig:lowZmodels}) with the solar--metallicity
models (figure~\ref{fig:models}).
Using Starburst99 and Cloudy, we find that Z$= 0.2$ times solar models have
initial mid--infrared line ratios that are $\sim 3$ times greater than 
solar--metallicity models; these line ratios fall more slowly with time
than in solar--metallicity models.

While metallicity affects the mid--infrared line ratios, metallicity 
can be measured and corrected for.
Within galaxy samples that have similar measured metallicity, 
uncertainties in the metallicity should affect the
mid--infrared line ratios by factors that are much smaller
than the orders--of--magnitude changes in line ratio values expected due to \Teff\ 
(as discussed in \S~\ref{sec:specsynth}).

{\bf Ionization parameter and morphology}. The ionization parameter, as the ratio of the 
spectral intensity to the gas density, combines two of the fundamental parameters that 
determine the degree of ionization in a nebula. In a Galactic \HII\ region, the 
ionization parameter changes rapidly with radius because of the 1/R$^2$ falloff and 
absorption of UV photons by the nebula (which also alters the spectral shape.)  
Morphology then determines which parts of the nebula influence others.  
A starburst galaxy, however, is much messier than an assembly of pseudo-spherical 
\HII\ regions: the ISM is generally fragmented, and gas parcels are ionized by 
many stars.  For example, in M82, it appears that the interstellar medium 
is highly fractionated (e.g., Seaquist, Frayer, \& Bell 1998) \nocite{seaquist}
and that the whole $\sim 450$~pc nuclear starburst and individual 
$\sim 20$~pc star--forming clusters can be described by a single ionization 
parameter (\citet{thornley}, citing the dissertation of N.~F\"{o}rster Schreiber.) 
Thus, it seems more appropriate to model a starburst as though the gas and stars 
are thoroughly mixed (by employing a mean UV spectrum and mean ionization parameter), 
rather than as a collection of spherical clouds, each with a single ionization source. 
This ``mixed gas and stars'' model is achieved in practice by assuming plane--parallel 
geometry and a composite ionizing spectrum.

Ionization parameters (U) have been estimated in several nearby
starburst galaxies by measuring the number of Lyman continuum
photons and the size of the starburst region.
\cite{thornley} summarize measurements in NGC~253, NGC~3256, and M82, 
which are all consistent with $\log U = -2.3$.
Measurements have also been obtained for Arp~299 \citep{arp299};
NGC~1614 \citep{ngc1614};  NGC~1808 \citep{ngc1808};
IC~342 \citep{ic342}; NGC~6946 \citep{ngc6946};
and NGC~3049 \citep{ngc6946}.  In addition, \citet{hbt} measure the 
Lyman continuum flux in fourteen nearby galaxies.
Six of these galaxies have multiple measurements of $\log U$, which gives
some estimate of the (often considerable) uncertainty.

\begin{figure*}
\figurenum{10}
\includegraphics[angle=270]{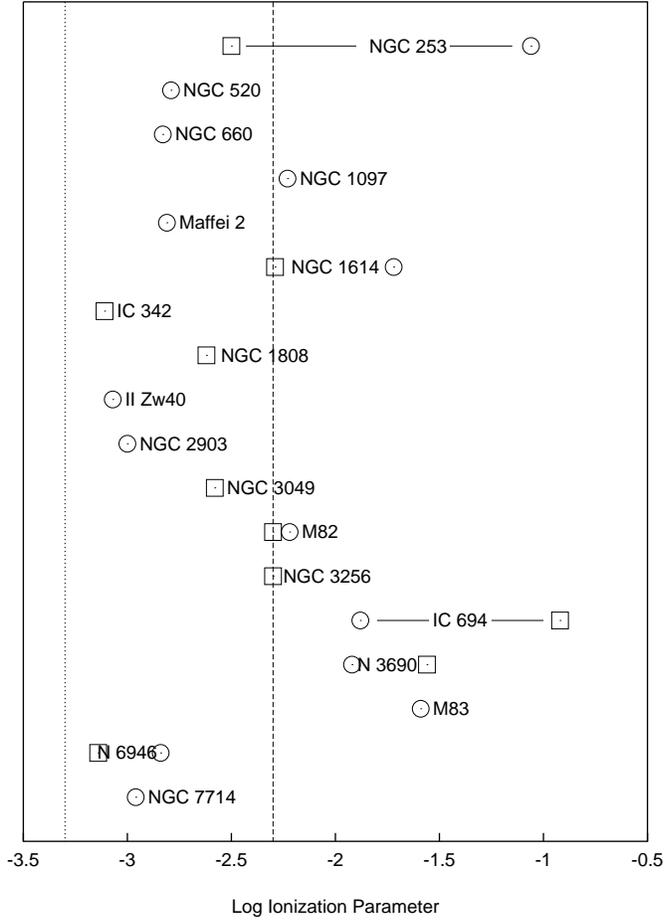}
\figcaption{Ionization parameters for local starburst galaxies, found using 
Lyman continuum fluxes and starburst radii from the literature
(see \S~\ref{sec:depends} for references.)  Data from \citet{hbt} are plotted as
circles; other data are plotted as squares.  The vertical dashed line shows
$\log U = -2.3$, the value we have adopted as typical for starbursts.  The
vertical dotted line is the value that would depress [\NeIII]/[\NeII] by
a factor of seven relative to our models, which is what is required to 
make the observed line ratios consistent with a \Mup $= 100$~\Msol\ 
Salpeter IMF.}
\label{fig:IP-local}
\end{figure*}

In figure~\ref{fig:IP-local}
we plot the ionization parameters derived from these studies.
It should be noted that each of these $U$ values is actually a 
\emph{lower limit}, since we use the maximum radius of the starburst region
to compute the ionization parameter.
When the gas density was not measured, we assume $n_e = 300$~\cc; 
the true ionization parameter scales as
$IP = IP_{300} - \log (n_e / 300~\cc)$.  
Figure~\ref{fig:IP-local} shows that the ionization parameter used 
in our simulations, $\log U \le -2.3$, is a reasonable average value given 
the measurements available for nearby starbursts.  

How sensitive are the mid--infrared line ratios to the ionization parameter?
Reducing $U$ in our models by a factor of $10$ lowers the 
[\NeIII]~$15.6$~\micron\ / [\NeII]~$12.8$~\micron\ line ratio by a factor 
of $\sim 7$.
Therefore, if the ionization parameters of starburst galaxies vary by a factor of 
$\sim10$ or more, this parameter could account for considerable spread in 
observed mid--infrared line ratios.
However, there is no tendency for galaxies with small [\NeIII]/[\NeII] to 
have low ionization parameters in figure~\ref{fig:IP-local}, indicating
that $U$ is not the dominant parameter determining this flux ratio.
Comparing with the restricted
range of {\it U} observed in starbursts, we conclude from the modeling in 
\S~\ref{sec:specsynth} that \Teff\ dominates variations in this line ratio 
in such regions.

Starburst ISM morphologies are far too complex to reproduce in simulations; fortunately, 
parameterization of a starburst by a single, global ionization parameter and a mean 
ionizing spectrum is physically motivated, agrees with observations, and simplifies the 
problem sufficiently to allow modeling.

Another test of the diagnostic usefulness of the mid--infrared line ratios is 
provided by studies of Galactic \HII\ regions.  \citet{mh2} found that, in compact 
\HII\ regions, the line ratios 
[\NeIII]~$15.6$~\micron / [\NeII]~$12.8$~\micron, 
[\ArIII]~$8.99$~\micron / [\ArII]~$6.99$~\micron, and
[\SIV]~$10.5$~\micron / [\SIII]~$18.7$~\micron\
correlate very well with each other, suggesting their reliability.
\citet{morisset-apj} has also demonstrated the use of these lines to 
estimate \Teff\ and $U$ in Galactic \HII\ regions, though as demonstrated
by \citet{morisset-aa}, outside constraints on ionization parameter and 
metallicity are usually necessary.

\subsection{The Spectrum of the Ionizing Radiation}

We now focus on using the fine structure line ratios to estimate the spectrum of the 
ionizing radiation in starbursts. As figure~\ref{fig:models} illustrates, once an 
instantaneous burst is older 
than $6$~Myr, [\NeIII]/[\NeII], [\SIV]/[\SIII], and [\SIV]/[\NeII] are so low ($<0.001$) 
that the higher--ionization line should not be detected. [\ArIII]/[\ArII]
and [\ArIII]/[\NeII] fall off more slowly, but still require a dynamic range
exceeding $100$ to detect both lines in each ratio.
Such very low line ratios are not seen in \citet{thornley}, which with $27$ 
[\NeIII]~$15.6$~\micron/[\NeII]~$12.8$~\micron\ 
measurements is the largest sample to date of mid--infrared fine structure 
lines in starburst galaxies.  The lowest ratio detected by \citet{thornley} is $0.05$, 
and $5$ galaxies have upper limits.  The simplest explanation of this behavior is that 
massive stars continue to form at low rates after the peak of a starburst.

In the \citet{thornley} sample, all but $3$ galaxies have [\NeIII]$<$[\NeII].
The three outliers, with neon ratios from $1$ to $12$, are all low--mass, 
low--metallicity galaxies (NGC~55, NGC~5253, and II~Zw~40).  We will consider the 
higher--metallicity galaxies now, in the context of the solar--metallicity models, 
and defer discussion of the low--mass, low--metallicity galaxies to 
\S~\ref{sec:lowZspecsynth}.

\begin{figure*}
\figurenum{11}
\plotone{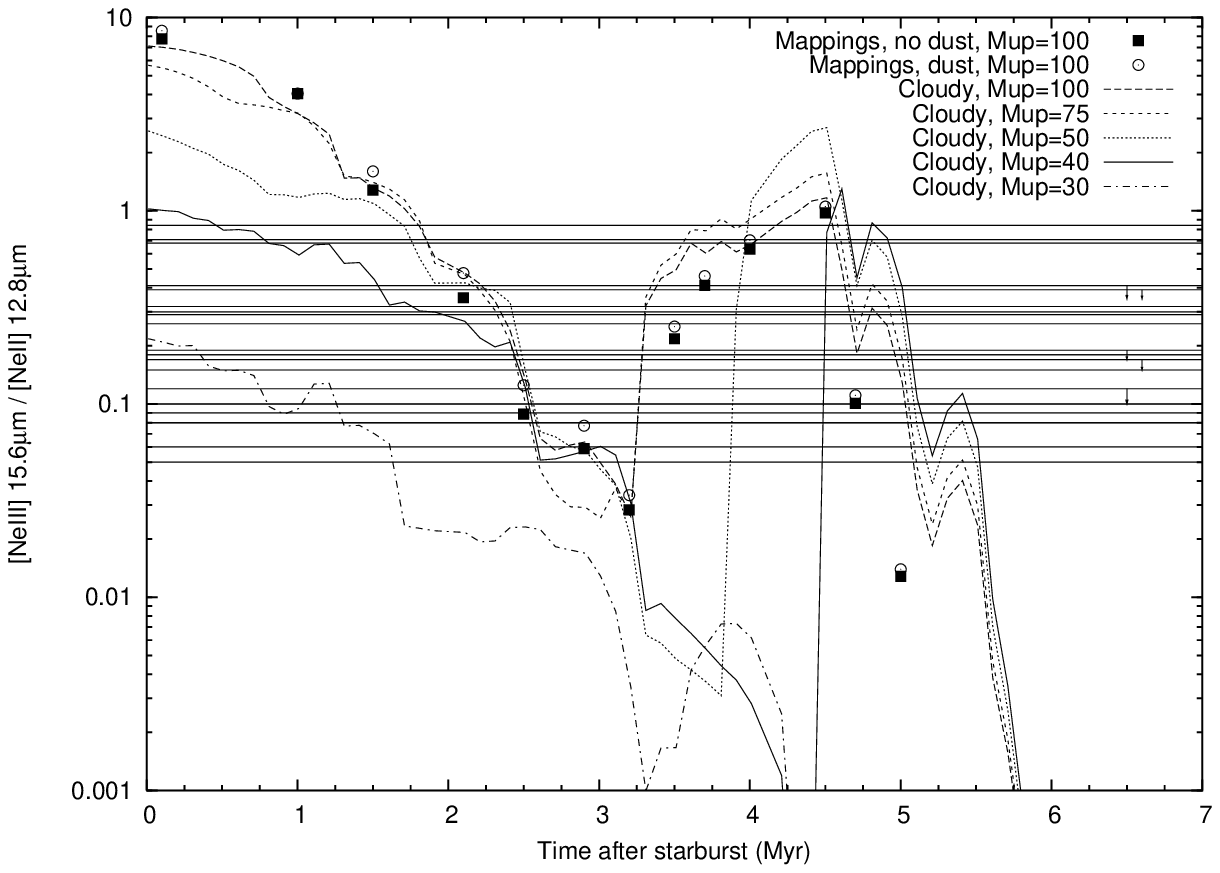}
\figcaption{
Burst models of [\NeIII]/[\NeII] for solar metallicity, and 
observed values.  The curves are Cloudy models with varying \Mup; squares 
show Starburst99/Mappings \Mup$=100$~\Msol\ models without dust; 
circles show dusty Starburst99/Mappings models with \Mup$=100$~\Msol.  
The \Mup$=100$~\Msol\ Cloudy and Mappings models use the same ionizing 
spectra, but different photoionization codes and treatment of the 
mid--infrared lines; the two photoionization codes are in generally good 
agreement.  Overplotted are 
the observed [\NeIII]/[\NeII] ratios of \citet{thornley}.  The
neon line ratios of the three low--metallicity galaxies II~Zw~40, NGC~5253, 
and NGC~55 are not plotted; they should be considered in light of 
the low--metallicity models of figure~\ref{fig:lowZmodels}, and are 
overplotted there.
}
\label{fig:ne-newplot}
\end{figure*}

In figure~\ref{fig:models}, as \Mup\ decreases, the line ratios 
decrease during the main sequence phase (because the ionizing spectrum softens), 
and the gap widens between the two phases of high line ratios (because fewer 
stars become Wolf--Rayets.)  We now consider these models in light of the 
measured neon ratios of \citet{thornley}, which are overplotted in 
figure~\ref{fig:ne-newplot}. 

In the \Mup$=100$~\Msol\ model, for $46\%$ of the first $5$~Myr, the predicted 
[\NeIII]/[\NeII] exceeds the highest line ratio measured by \citet{thornley} 
for a high--mass, $\sim$solar--metallicity galaxy; thus, this model poorly fits
the data.  A much better fit is the $Z=Z_{\sun}$, \Mup$=40$~\Msol\ model.  
For only $6\%$ of the first $5$~Myr does this model predict [\NeIII]/[\NeII]$>1$; 
for $65\%$ of that time, it predicts neon line ratios within the range of the 
Thornley detections.  The \Mup$=40$~\Msol\ model fits markedly better than the 
\Mup$=50$ and $30$~\Msol\ models.  Because one--quarter of the Thornley datapoints
are upper limits (excluding the three low--mass, low--metallicity galaxies), 
the M$_{up}=40$ model is a better fit to the Thornley data than the above 
percentages indicate.

\begin{figure*}
\figurenum{12}
\plotone{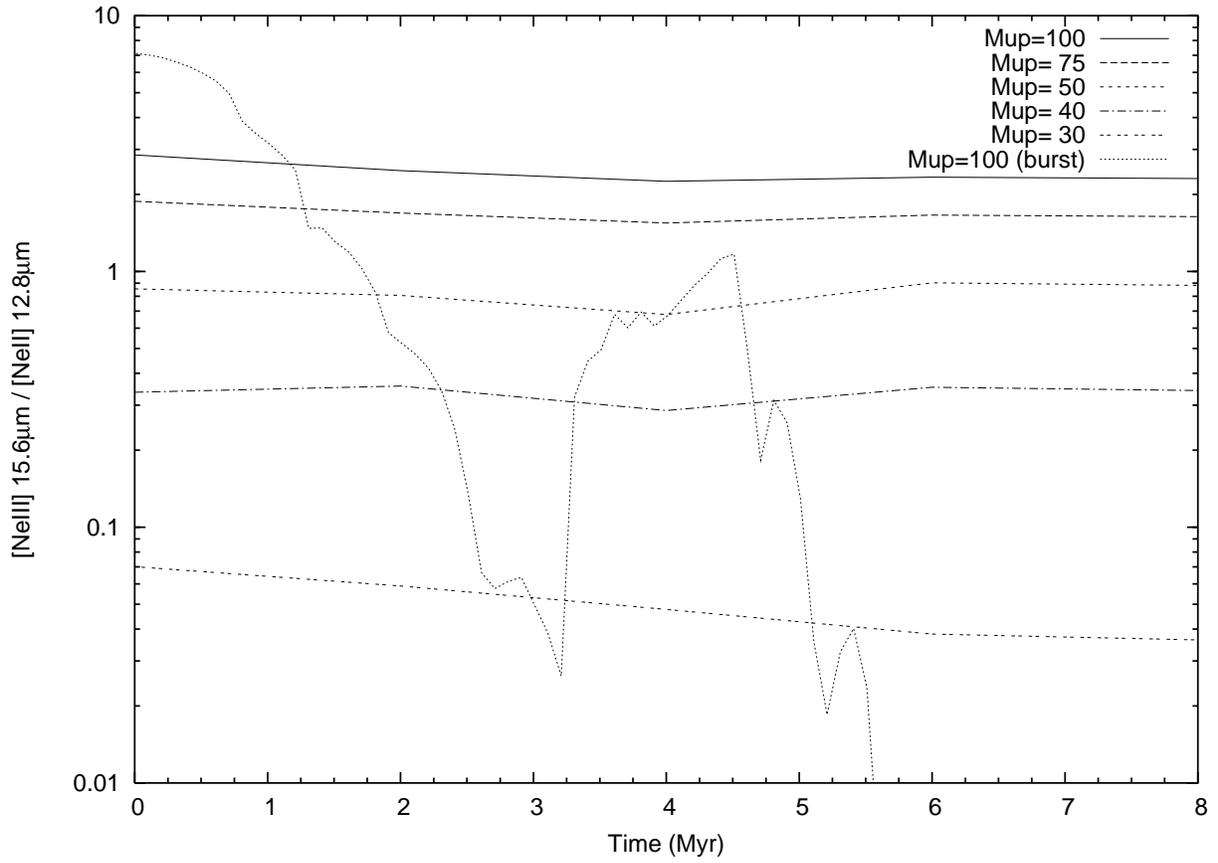}
\figcaption{Continuous star-formation models of [\NeIII]/[\NeII] 
for solar metallicity.  Five differerent \Mup\ values are plotted.  For 
reference, we overplot (finely dotted line) the solar--metallicity, 
\Mup$=100$~\Msol, instantaneous burst model from figure~\ref{fig:models}.
}
\label{fig:continuous}
\end{figure*}

One draws the same conclusion from continuous star formation models, as shown in
figure~\ref{fig:continuous}.  Such models with \Mup$=100$ and 
$75$~\Msol\ predict a constant neon ratio above $1$,
while the neon ratio for the \Mup$=30$~\Msol\ model falls below the Thornley range.  
The \Mup$=40$ and $50$~\Msol\ models predict neon line ratios within the Thornley range;
the \Mup$=40$ model comes closer to the median.

These results are consistent with those of \S~\ref{sec:midir-test}, which found that
the heteronuclear and homonuclear mid--infrared 
line ratios within four regions of M82 required \Mup$<50$ to  \Mup$<65$~\Msol\
(depending on the region), that He~2--10 required \Mup$<65$~\Msol, and that
NGC~253 required \Mup$<100$~\Msol.
Thus, [\NeIII]/[\NeII] in the high--mass, solar--metallicity \citet{thornley} galaxies,
and a concordance of line ratios in M82 and He~2--10, are all significantly lower than 
the predictions of a Salpeter IMF extending to $100$~\Msol.  An IMF that
is deficient in massive ($\ga40$~\Msol) stars could produce the observed 
line ratios.  

\subsection{Ionizing Conditions in Low Metallicity Starbursts}
\label{sec:lowZspecsynth}
We now discuss mid--infrared line ratios in low metallicity starbursts.
Lowering the metallicity from solar elevates the mid--infrared line ratios, and fills
in the valley between the MS and WR phases. 
\HeIh/\Brten\ is completely saturated until the WR stars die, by contrast to 
its double--peaked behavior for solar metallicity.  
Overplotted in figure~\ref{fig:lowZmodels}  
are the line ratios for II~Zw~40, NGC~5253, and NGC~55, in order of 
decreasing [\NeIII]/[\NeII] ratio from \citet{thornley}.  
All three of these galaxies have low metallicity:
II~Zw~40 has measured [O/H] $=0.20\pm0.01$ \citep{diazetal} and 
[O/H]$=0.19\pm0.04$ \citep{garnett89}, [S/H]$=0.12\pm0.03$ \citep{garnett89}, 
and [Ne/H]$=0.3$ \citep{martinhernandez}, all linear and relative to solar abundance.
NGC~5253 has measurements of [O/H]$=0.28$ 
(Storchi--Bergmann, Kinney, \& Challis 1995)
\nocite{sb} 
and [Ne/H]$=0.58$ \citep{martinhernandez}.  
NGC~55 has measured [O/H]$=0.25$ to $0.37$ \citep{websmith}.

We consider the line ratios of these galaxies in light of the 
low--metallicity models.  For NGC~55, the only mid--infrared line ratio 
available in the literature is [\NeIII]/[\NeII];  the observed value can 
easily be produced by any \Mup\ from $30$ to $100$~\Msol.
For II~Zw~40, the observed [\NeIII]/[\NeII]$=12$ cannot be achieved by Starburst99/Cloudy
models with solar metallicity.  With the low metallicity models, we find that the
observed [\NeIII]/[\NeII], [\SIV]/[\NeII], and [\ArIII]/[\NeII] line ratios
cannot be produced at any age unless \Mup\ is greater than $40$~\Msol.
The \HeIh/\Brten\ ratio agrees that the ionizing field is rather hard, but 
is insensitive to \Mup.
For NGC~5253, unless the burst is $<0.5$~Myr old, the measured [\SIV]/[\NeII] 
requires \Mup$>40$~\Msol.  This constraint is strengthened if we consider the 
[\NeIII]/[\NeII], [\ArIII]/[\NeII], and \HeIh/\Brten\ ratios, which all predict 
ages within $3$ to $5$~Myr, for a broad range of \Mup\ ($40$ to $100$.)  
The \HeIh/\Brten\ constraint is particularly  insensitive to \Mup.  
If one assumes this age range, the [\SIV]/[\NeII] ratio requires
\Mup$>60$~\Msol. 
 
Thus, while high--mass, solar--metallicity starburst galaxies are seen to have
lower [\NeIII]/[\NeII] ratios than a Salpeter IMF with \Mup$=100$~\Msol\
predicts, the low--metallicity galaxies 
II~Zw~40 and NGC~5253 have the high neon ratios expected if they contain very 
massive stars.

\subsection{Caveats and Assumptions}
\label{sec:caveats}
How robust is the conclusion that the nebular line ratios indicate that most 
high--mass, solar--metallicity starbursts have soft ionizing continua?
First, we have assumed that the Thornley galaxies are generally of solar metallicity.  
If they were more metal--poor, this would raise the predicted line ratio curves, 
and thus increase the discrepancy between the predicted and observed ratios.
The opposite effect (weakening our constraint) occurs if the 
Thornley galaxies have super--solar metallicity.  \citet{thornley} use the 
strong--line method to derive metallicities
of $1.9\pm 1$~Z$_{\odot}$ for $13$ of their galaxies (excluding NGC~5253 and II~Zw~40.)
This result is consistent with the metallicities from optical line ratios, but we 
prefer the Thornley mid--infrared estimate because it should be reddening-independent.
 Starburst99 is not optimized for such metallicities, but we use twice--solar models 
nonetheless to crudely estimate whether super--solar metallicities could void our result.
For \Mup$=100$, $75$, and $60$~\Msol, doubling the metallicity from solar lowers the 
[\NeIII]/[\NeII] line ratios and increases the duration of the WR phase by 
$\sim0.5$~Myr, which brings the models closer to agreement with observations, 
but deepens the trough between the main sequence and WR phases to $100\times$ 
below the lowest Thornley detection.
These models predict neon line ratios within the observed Thornley range for $\sim40\%$
of the first 6~Myr---little better than the solar--metallicity \Mup$=100$ model.
To summarize, while uncertainties remain because metal--rich stellar evolution is not 
well understood, current models indicate that the low line ratios observed  
in starburst galaxies are unlikely to be explained away by metallicity effects.

Another way to negate the conclusion would be for the high--mass, 
solar--metallicity starburst galaxies to have much lower ionization 
parameters than we assumed.  For the observed neon line ratios to arise in 
starbursts with \Mup$=100$~\Msol, the starbursts must have $U$ 
about $10$ times weaker than our assumed $\log U_{max}=-2.3$.  None
of the 18 galaxies in figure~\ref{fig:IP-local} has a measured 
ionization parameter this low.

In fact, because in our models the ionization parameter starts at $\log U = -2.3$
and falls with the ionizing flux, the ionization parameter in our models is
already fairly low.  (For example, 5~Myr after
a solar--metallicity, \Mup$ = 100$~\Msol\ burst, the ionization parameter has 
fallen to $\log U = -3.15$.)  Thus, our ionization parameter assumptions are 
conservative in that they tend to predict low line ratios for a given \Mup;
as a result, when comparing to observed line ratios, the models will
be slightly biased toward finding high \Mup.  Thus, the choice of ionization
parameter is not the reason we find generally low \Mup\ in starburst galaxies;
the models are actually biased against finding this result.

For simplicity, we have modelled star formation as an instantaneous burst.
Starburst galaxies are of course more complicated.  An instantaneous burst is 
the \emph{most conservative} assumption of star formation history for
the purpose of constraining \Mup.  As illustrated in figure~\ref{fig:continuous},
extended star formation or a series of bursts would elevate predicted line ratios 
above the instantaneous--burst case for most of the burst duration.  As such, 
extended star formation would increase the discrepancy between the low ratios
observed in starbursts and the high ratios predicted by high--\Mup\ models.  

Dust grains harden the ultraviolet ionizing continuum, as pointed out by 
\citet{aannestad}.  Thus, if dust competes for the ionizing photons, this
elevates the line ratios, and our conclusions are strengthened.  
Figure~\ref{fig:ne-newplot} shows this effect in Starburst99/Mappings models
with and without dust.  These models were created using the Starburst99/Mappings~III 
web interface, beta test version 3q \citep{mappings}.  That figure also shows that
the two different photoionization codes Mappings and Cloudy, given the same input 
spectra and nebular conditions, predict very similar neon line ratios.  This helps 
address the concern that our results depend on the reliability of photoionization 
codes and their input atomic constants.

The other major assumptions in our work are the choice of 
stellar evolution tracks and hot stellar spectra.  Had we used 
the (hard--spectrum) CoStar models, they would have increased the predicted line
ratios and made the \citet{thornley} galaxies seem even more deficient
in high--mass stars.  Thus, our use of the softer \citet{pauldrach01} 
atlas is conservative in terms of existing hot star models.  However, our 
conclusions could be invalidated if real stars have much softer ionizing 
continua than \citet{pauldrach01}.

We note that NGC~3077, 4214, and 4861 now have well--measured, saturated 
\HeIh/\Brten, but no published mid--infrared spectra.  Mid--infrared
spectra of these galaxies should further test the trends in nebular line behavior 
discussed in this paper (all of these galaxies would appear to fall into the 
low--mass, low--metallicity category). 

\subsection{UV and Nebular Diagnostics in Conflict?}
\label{sec:UV}
The very massive stellar populations of a number of starburst galaxies 
have been constrained by ultraviolet spectroscopy. In cases where the 
burst age is more than $\sim5$~Myr, the UV spectra cannot test for stars 
above $40$~\Msol\ because the most massive stars have already exploded 
as supernovae or evolved off the main sequence (e.g., \citet{delgado99}).
A small number of starbursts have strong P~Cygni profiles indicative of a 
very young burst and the presence of very massive stars. 
Thus, there appears to be a tendency for ultraviolet 
spectra of stellar populations to indicate larger \Mup\ than do the nebular 
lines (although the galaxy samples observed in the UV 
and mid--infrared hardly overlap).   
We now consider the cases of He~2--10 and NGC~3049; the UV 
spectra of both these starburst galaxies show P~Cygni profiles, and nebular spectra
are available (mid--infrared for the former galaxy, and optical for the latter.)

He~2--10 is an extremely rare case of a starburst which has available ultraviolet 
spectra of adequate quality to search for P~Cygni line profiles as well as 
high--quality mid--infrared line measurements. 
Although He~2--10 is of low mass and metallicity globally, the abundances in its 
nuclear \HII\ regions are approximately solar (Kobulnicky, Kennicutt, \& Pizagno 1999).
\nocite{chip}
Best fits to the UV spectrum require \Mup$\ge 60$~\Msol\ \citep{chandar}.
From our modelling of the mid--infrared line ratios, we find \Mup$<65$~\Msol.
Thus, these observations permit a discrepancy between the diagnostics, 
but do not require one.

For NGC~3049, mid--infrared spectra are not available, but optical and UV spectra are.
This galaxy is of solar (or slightly higher) metallicity in the starburst regions
(Guseva, Izotov, \& Thuan 2000)  \nocite{guseva}
although it is of low mass and luminosity, and hence probably of 
low global metallicity.  
\citet{delgado} find that the P~Cygni line profiles of \CIV\ and \SiIV\ 
in NGC~3049 require \Mup$\ge60$~\Msol, and rule out ages younger than $2.5$~Myr
and older than $4$~Myr.
Further, they find that the UV diagnostics disagree with optical nebular lines 
as to whether massive stars are present; they fit the optical nebular
lines by a \Mup$=40$~\Msol, t$=2.5$~Myr model---parameters which would not
create the observed P~Cygni profiles in the UV.  Given these results, 
the authors question whether nebular line ratios can reliably indicate the 
presence of massive stars.

We therefore re--examine the nebular line results for NGC~3049.  
In modelling these lines, \citet{delgado} used an older 
version of Starburst99 that employed pure helium WR models 
(Schmutz, Leitherer, \& Gruenwald 1992) \nocite{schmutz} 
and Kurucz O~star atmospheres prepared by Lejeune, Cuisinier, \& Buser (1997).
\nocite{lejeune} 
An update of Starburst99 incorporating new stellar models 
(\citet{pauldrach01} and \citet{hillmill}, as packaged by \citet{snc}) 
became available after submission of their paper.  
The authors note that these new stellar models would soften the ionizing 
spectrum and reduce the discrepancy with the UV results, but they
did not make a detailed reconciliation.

Using our models, which make use of these new stellar atmospheres, we 
re--examine the nebular lines of NGC~3049.  In \citet{delgado}, 
$\log U$ is fixed with time, and varies with radius ($\log U = -2.58$ at 
R$_{max}=100$ pc) in a spherical model.  This results in a generally
stronger $U$ than in our models, in which $U$ falls with time.
To compare with the results of \citet{delgado}, we ran new models with 
$\log U$ fixed at $-2.3$.  This value for $\log U$ is within the measurement 
uncertainties of n$_H$ and Q(H) of the value used by \citet{delgado}.
This choice of slightly higher ionization parameter 
biases our test toward low values of \Mup\ (and agreement with the results of 
\citet{delgado}).

We consider the age range $3\le t \le 4$~Myr, as required by the UV lines \citep{delgado}.  
Over this time period, 
\HeI~5876/\Hbeta\ can be fit by $40< \Mup \le 100$~\Msol, and
[\OIII]~5007/\Hbeta\ by  $40 \le \Mup \le 100$~\Msol.
[\NII]~6584/\Hbeta\ only requires \Mup$<75$, 
[\SII]~6716/\Hbeta\ and [\OII]~3727/\Hbeta\ can be fit by any \Mup\ from $30$ to $100$,
and  [\SII]~6731/\Hbeta\ and [\OI]~6300/\Hbeta\ cannot be fit by any model.
Thus, even using a high $U$ model, we do not find that low \Mup\ is required.
The updated stellar models remove the inconsistency between the UV
and nebular lines noted by \citet{delgado} in NGC~3049.

\section{DISCUSSION}
We have used a number of tests to show that the nebular line ratio 
[\NeIII]~$15.6$~\micron/[\NeII]~$12.8$~\micron\ is a robust measure of the 
hot stellar population in starbursts. The line ratio is virtually unaffected 
by extinction, and as a homonuclear ratio involving a rare gas it is not subject 
to abundance variation or depletion onto dust.  Where it can be compared to other 
reliable \Teff\ indicators, the agreement is good.  Since the mid--infrared neon 
lines vary over several orders of magnitude during a few million years of starburst 
evolution, measurements of moderate precision can give good \Teff\ constraints. 

The neon ratio indicates low \Teff\ in all members of a reasonably large sample of 
massive, high--metallicity starburst galaxies \citep{thornley}. We have shown 
that plausible modifications to the interpretive models (adding dust, lowering the 
metallicity, changing the ionization parameter) leave the basic constraint of low 
\Teff\ unchanged or strengthened. 

The conclusion from [\NeIII]~$15.6$~\micron/[\NeII]~$12.8$~\micron\ contrasts 
with the evidence for massive, hot stars from P~Cygni line 
profiles in the ultraviolet spectra of two galaxies, He~2--10 and NGC~3049. 
In the first case, we find that the infrared nebular lines are consistent with 
the hot stellar spectrum indicated in the UV. In the second case, no 
mid--infrared spectrum exists, and the optical spectrum of \citet{delgado} does not 
conflict with the UV result. Thus, there is no overt conflict between the P~Cygni 
lines in UV starburst spectra and the limits on \Teff\ set by nebular lines.  
However, to account for the observed low--excitation nebular spectra of starbursts,
galaxies like NGC~3049 must represent a very rare stage in
starburst evolution.  Can the UV wind observations, infrared nebular line results, 
and starburst models be reconciled, given this new constraint?

Our calculations of the emission--line properties of starbursts are based on 
traditional synthesis modelling, as introduced by \citet{rieke1980}. 
Such modelling makes the assumption that newly--formed stars appear on the main 
sequence according to an assumed formation rate with masses given by an 
initial mass function. It has recently become popular to assume a Salpeter IMF, 
although \citet{rieke-m82} derived a very similar IMF {\it ab initio} to fit 
the starburst properties of M82.  (Both of these IMFs differ significantly from 
estimates of the local IMF, in that both have a substantially larger portion of 
massive stars).  

Assuming a Salpeter IMF extending to $100$~\Msol, we have shown that these 
models predict an early phase in starburst evolution, of duration $3$ to $4$ 
million years, when hot, massive stars should produce
high--excitation emission lines.  ``Starbursts'' are identified as--such 
up to ages of $15$ to $20$~Myr; thus, about $20\%$ of active starbursts should
be in the early $<4$~Myr phase.
  However, the data of \citet{thornley} show no starbursts 
in massive, high--metallicity galaxies with the line ratios predicted for this 
early phase.  One explanation for this discrepancy would be that the Salpeter 
IMF substantially overestimates the numbers of very massive stars.  We have 
shown that the mid--infrared line ratios can be explained if the IMF cuts 
off at $40$ to $50$~\Msol.  Parameterizing the IMF by a cutoff is an 
oversimplification; a substantial steepening of the IMF slope is probably
a more appropriate description.  One advantage of such an IMF is that it 
suppresses the production of oxygen, which can otherwise reach very high 
abundances in starbursts \citep{rieke-m82}.

However, in addition to the indications from UV spectra that stars more massive 
than $40$~\Msol\ can form in substantial numbers in starbursts, the Arches Cluster
near the center of the Milky Way has a large population of $\sim100$~\Msol\ 
stars \citep{figer}.  (The mid--infrared line ratios in the Arches \citep{giveon}
are consistent with a burst of age $2$--$3$ or $6$~Myr in our models, assuming 
twice--solar metallicity.)
None of these observations can confirm the standard assumption of a Salpeter 
IMF extending to $100$~\Msol, and the possibility of rolloff in the IMF toward 
very high masses needs to be considered in detail.  However, the Arches
and the UV starburst results suggest it is unlikely 
that the lack of high--excitation emission lines can be explained entirely in 
terms of a substantial steepening in the IMF above $40$ -- $50$~\Msol.

We have therefore searched for other causes for this behavior.  We believe 
an explanation can be found in an incorrect assumption in the standard synthesis 
models:  that the full luminous output of newly--formed stars escapes into 
surrounding diffuse gas.  This assumption justifies modelling starbursts as
traditional low--density \HII\ regions.  Instead, we suggest that the majority
of massive stars in starbursts spend a substantial part of their main
sequence lifetimes embedded within dense, highly--extincted regions---similar to
the ultracompact \HII\ regions of the Milky Way---and are thus
invisible to optical, near--infrared, and mid--infrared nebular line studies. 

In the solar neighborhood, it appears that about $15\%$ of the main sequence life 
of a massive star is spent within an ultracompact \HII\ region 
(Kurtz, Churchwell, \& Wood 1994).  \nocite{kurtz}
Hanson, Luhman, \& Rieke (1996) have detected \nocite{hanson}
in the near--infrared about half of a sample of radio--selected ultracompact 
\HII\ regions.  They
conclude that the detected regions typically are obscured by A$_V$ $= 30$ -- $50$. 
Since the undetected regions in their sample should be even more heavily obscured, 
we take a typical case to be A$_V$ $\sim 50$.  Thus, these objects would not
contribute to the optical or near--infrared emission--line spectra of the Milky Way.
The heavy extinction would even diminish the fluxes of the mid--infrared fine 
structure lines such as [\NeII]~$12.8$~\micron\ and [\NeIII]~$15.6$~\micron\
by a magnitude or more.
More importantly, the densities in many ultracompact \HII\ regions exceed 
the critical densities for these lines (e.g., $2 \times 10^5$~cm$^{-3}$
for [\NeIII]~$15.6$~\micron).  Thus, even in the solar neighborhood, 
the accuracy of traditional synthesis models would be improved by assuming that 
massive stars contribute their bolometric luminosity to the region for their 
entire main sequence lifetimes, but influence the usual \Teff\ indicators in 
emission line spectra for only $85\%$ of their lives. 

The correction suggested above would be small for synthesis modelling of the 
solar neighborhood.  However, if the ultracompact \HII\ region lifetimes 
were significantly 
greater, a substantial deviation from traditional synthesis models would be expected. 
For nuclear starbursts in massive galaxies, the external pressure is large, due to 
both the high density and high temperature of the interstellar medium.  As a result, 
the ultracompact \HII\ regions of starbursts should be small and their 
expansion retarded compared with Galactic ones 
(De Pree, Rodr\'iguez, \& Goss 1995; Garci\'a-Segura \& Franco 1996).
\nocite{depree}  \nocite{garcia}
The gravitational field of the central star(s) should also play an important role, 
slowing the expansion further \citep{keto}.   
Thus, it is likely that the massive stars in nuclear starbursts spend a substantial 
fraction of their lifetimes embedded in high--extinction regions.  It is even
plausible that this phase is only terminated when these stars begin 
to lose mass in strong winds---the evolutionary phase seen in UV spectra of 
starbursts.  This possibility is suggested by the failure, to date, to detect 
any nuclear starburst that appears younger than 
about 3 million years, based on either nebular line ratios or UV spectroscopy.  

Another indication supporting the UC\HII\ hypothesis is that 
starburst models under--predict the observed bolometric
luminosities of starbursts.  Further evidence
is that radio recombination lines and free--free continua in starbursts 
indicate substantially more extinction than indicated by the Brackett lines.
For example, \citet{chad} deduced A$_V$ $\sim 50$ to the ionized gas in 
NGC~253 and suggested that much of this gas lies in very compact \HII\ regions. 

In addition, Beck, Turner, \& Kovo (2000) \nocite{btk}
found a substantial population of sources in young starburst galaxies whose spectra
\emph{rise} from $\lambda=6$~cm to $\lambda=2$~cm, indicating self--absorbed
(optically thick to electron scattering) emission.  Parsec--scale estimated 
sizes and large ionizing fluxes (estimated at roughly $10^3$ to $10^4$ OB stars)
suggest that these sources are highly obscured young superstar clusters---similar
to ultracompact \HII\ regions, but containing many more stars.
Ironically, these results apply to low--metallicity,
low--luminosity galaxies similar to those that emit high--excitation nebular lines.
The other galaxies known to have non-AGN rising spectrum sources are 
NGC~5253 (Turner, Ho, \& Beck 1998) \nocite{thb} and He~2--10 
(Kobulnicky \& Johnson 1999; Vacca, Johnson, \& Conti 2002).
\nocite{kj99,vjc}
Massive spiral starburst galaxies do not show this self--absorbed emission, though 
it is plausible that such sources exist but are are hidden by non-thermal 
emission from supernovae.

There are a number of consequences for starburst modelling.  Because the duration 
of the obscured phase may depend on the mass of the central star, it may be difficult 
to deduce an accurate IMF in starbursts using nebular diagnostics. The derivation 
of the IMF from fitting UV spectra would also be suspect, since there could be a 
mass/age dependence on the stars contributing to these spectra rather than their 
providing a snapshot of the integrated hot stellar population.  In addition, by 
suppressing the signatures of the youngest stars, the UC\HII\ stage will tend to 
make the duration of starbursts appear artificially short. 

\section{CONCLUSION}
\label{sec:summary}
We have obtained $1.7$~\micron\ and $2.1$~\micron\ spectra for six
nearby circumnuclear starburst galaxies to measure the \HeIh/\Brten\
and \HeIk/\Brgam\ line ratios.
Simple recombination physics and independence from nebular conditions
and extinction make \HeIh/\Brten\ an accurate diagnostic of the hardness of the 
ionizing continuum (``\Teff'').
The lines are too weak to be used in distant galaxies,
but in nearby galaxies can test more detectable but potentially 
problematic diagnostics.

We present models for the behavior of the \Teff\ diagnostics discussed 
in this paper.  SEDs were generated by the spectral synthesis code 
Starburst99, which uses the most current O~star and Wolf--Rayet model spectra 
\citep{pauldrach01,hillmill}.
Although T$_{eff}$ remains a useful shorthand term, accurate models of the ionizing 
continuum must consider the entire population of hot main sequence and 
Wolf--Rayet stars.  Wolf--Rayet stars maintain high line ratio values 
after the O~stars have left the main sequence. Our updated models may aid 
interpretation of IRS spectra from SIRTF. 

We then test whether the observed line ratios are consistent with these new models.
By comparison with \HeIh/\Brten, we confirm that \HeIk/\Brgam\ is a 
problem--ridden diagnostic, as predicted \citep{shields}.  \HeIk/\Brgam\
also fails to correlate with the mid--infrared diagnostic [\NeIII]/[\NeII].
However, we point out that a low \HeIk/\Brgam\ ratio may indicate
a soft UV continuum.  NGC~4102, in which \HeIh/\Brten\ and 
\HeIk/\Brgam\ are both low, is an example.

We test [\OIII]~5007~\AA/\Hbeta\ as a \Teff\ diagnostic in starburst galaxies.
In comparison to \Heopta/\Halpha, [\NeIII]/[\NeII], and \HeIh/\Brten,
we find that [\OIII]/\Hbeta\ is systematically elevated. 
While aperture mismatch may contribute to the poor
correlation with [\NeIII]/[\NeII], the other two diagnostics were observed with
apertures comparable to that for [\OIII]/\Hbeta.  We suggest that 
shock--excitation of [\OIII]
by supernovae is the likely cause of very high [\OIII]/\Hbeta, and that 
this effect plus differing sensitivities to extinction may explain the considerable
scatter and lack of correlation with the other \Teff\ diagnostics.

We attempt to test optical He and H recombination line ratios.  Sample sizes 
in the literature are too small to compare \Heopta/\Halpha\ or 
\Heoptc/\Hbeta\ to each other or to [\NeIII]/[\NeII], \HeIh/\Brten, or 
\HeIk/\Brgam.  Instead, we estimate \Teff\ in
galaxies by the latter three diagnostics, and test for a correlation with 
\Heopta/\Halpha.  We find a correlation at $2\sigma$ significance. We re-evaluate the 
optical nebular lines of NGC~3049, and find them consistent 
with the age and IMF inferred from the UV spectrum; this removes
the discrepancy between the UV and nebular diagnostics noted previously using
less sophisticated stellar atmospheres.

We show that the mid--infrared lines
and \HeIh/\Brten\ give consistent answers for well-studied starbursts, lending 
credibility to the mid--infrared lines' use as diagnostics of starburst 
ionizing fields.  We also demonstrate theoretically that the behavior of these 
lines in starbursts should be dominated by T$_{eff}$ for galaxies with
similar metallicity. 

Having found the mid--infrared line ratios to be credible \Teff\ diagnostics, 
we use them to address the conditions in starbursts.   \citet{thornley} found 
low values of the [\NeIII]~$15.6$~\micron\ / [\NeII]~$12.8$~\micron\ ratio
in their sample of $27$ starburst galaxies.  In the context 
of our models, this result would suggest that high--mass, solar--metallicity 
starbursts form fewer M$\ga40$~\Msol\ stars than a Salpeter IMF.  Adding dust, 
lowering the metallicity, choosing a different stellar atlas, or choosing a more
extended star formation history would strengthen this conclusion.

However, another more likely possibility can account for this result. The relatively 
high density and temperature of the interstellar medium in nuclear starbursts 
should increase the duration of the ultracompact \HII\ region phase. As a result, 
most of the very massive stars may spend virtually their entire main sequence 
lifetimes embedded within dense, highly extincted regions, and thus will be nearly 
undetectable to conventional optical or near-to-mid-- infrared spectroscopy.
This situation will make it difficult to determine the high--mass IMF in starbursts. 

In contrast to the high--mass, solar--metallicity starbursts, in the low--mass, 
low--metallicity galaxies II~Zw~40 and NGC~5253, high neon line ratios seem to 
require stars more massive than $\sim40$--$60$~\Msol.  
This contrast can be 
understood if these galaxies form stars in regions where the interstellar medium 
is less effective at confining ultracompact \HII\ regions, or if the 
lifetimes of these regions are reduced at low metallicity.



\acknowledgments

We thank the Steward Observatory TAC for time allocation, telescope 
operator Dennis Means and the SO Kitt Peak staff, and Chad Engelbracht for
his FSPEC--specific \texttt{iraf} scripts.
We thank Gary Ferland for making Cloudy available to the astronomical
community, and Claus Leitherer for making Starburst99 available. 
We also thank Doug Kelly and Lisa Kewley for modelling 
advice, and Luis Ho for assistance using his atlas.  Don McCarthy, 
Ann Zabludoff, Ed Olszewski, and Dave Arnett provided comments that 
improved this paper. An anonymous referee provided an
exceptionally helpful critique of the original version of this paper. 
JRR was partially supported by an NSF Graduate Research Fellowship.  
This work was also partially supported by the MIPS Project, under
contract to the Jet Propulsion Laboratory.
This research has made use of the NASA/IPAC Extragalactic 
Database (NED) which is operated by the Jet Propulsion 
Laboratory, California Institute of Technology, under 
contract with the National Aeronautics and Space Administration.
This research has made use of the SIMBAD database, operated at CDS, 
Strasbourg, France.




\nocite{ccm}

\clearpage
\begin{table}
\tablenum{1}
\label{tab:c17}
\begin{center}
\caption{Dependence of C$_{1.7}$ on Electron Temperature (from \citet{vanzi})}
\begin{tabular}{ll}
\tableline\tableline
T$_e$ (K)  & C$_{1.7}$ \\
\tableline
5000   &  0.95  \\ 
10000  &  1.000 \\
20000  &  1.05  \\
\tableline
\end{tabular}
\end{center}
\end{table}

\begin{table}
\tablenum{2}
\label{tab:obs}
\begin{center}
\caption{Log of Observations for Circumnuclear Starburst Galaxies}
\begin{tabular}{lccc}
\tableline\tableline
\multicolumn{4}{c}{Integration times (seconds)} \\
Source	&  @$1.7$\micron & @$2.08$\micron & @$2.15$\micron \\
\tableline
\objectname[He 2-010]{He 2-010}   &  3840  &   \nodata & \nodata \\
\objectname[NGC 3077]{NGC 3077}   &  5760  &     960   &   960   \\
\objectname[NGC 3504]{NGC 3504}   &  5760  &     480   &   480   \\
\objectname[NGC 4102]{NGC 4102}   &  6720  &     960   &   960   \\
\objectname[NGC 4214]{NGC 4214}	&    1920  &       480 &   480  \\
\objectname[NGC 4861]{NGC 4861}	&    5760   &    \nodata & \nodata  \\
\objectname[NGC 6217]{NGC 6217}	 &   \nodata  &    960  &  960  \\
\tableline
\end{tabular}

\end{center}
\end{table}

\begin{table}
\tablenum{3}
\label{tab:waves}
\begin{center}
\caption{Vacuum Line Wavelengths\label{tab_wave}}
\begin{tabular}{ll}
\tableline\tableline
Transition & $\lambda$~(\micron)\\
\tableline
Br11   &  $1.6811$  \\
\HeI   &  $1.7007$  \\
Br10   &  $1.7367$  \\
\HeI   &  $2.0587$  \\
H$_2$  &  $2.1218$ \\
\Brgam &  $2.1661$ \\
\tableline
\end{tabular}

%
%
\end{center}
\end{table}

\begin{deluxetable}{lllclc}
\rotate
\tabletypesize{\small}
\tablecolumns{6}
\tablewidth{0pc}
\tablenum{4}
\label{tab:line_rats}
\tablecaption{Measured Line Ratios of Starburst Nuclei.\label{tab-line-rats}}
\tablehead{
\colhead{Object} & \colhead{$\HeIh/\Brten$} & \colhead{$\HeIh/\Brten$} & 
\colhead{$\HeIk/\Brgam$} & \colhead{$\HeIk/\Brgam$} & 
\colhead{$\HeIk/\Brgam$} \\
\colhead{source:} & \colhead{continuum--subtracted} & \colhead{raw, this work} & \colhead{this work}
& \colhead{literature} & \colhead{weighted mean}
}
%
\startdata
He~2--10  & $0.27 \pm 0.013$ & $0.23\pm0.05$ & \nodata & $0.52 \pm 0.03$ (1); $0.64 \pm 0.09$ (2) & $0.53 \pm 0.03$ \\
NGC~3077 & $0.46 \pm  0.085$ & $0.49\pm0.1$  & $0.49\pm0.07$  & $0.59 \pm 0.01$ (1) & $0.54 \pm 0.05$ \\  
NGC~3504 & $0.23 \pm  0.065$  & \nodata\tablenotemark{a}  & $0.49\pm0.05$  & $0.27 \pm 0.05$ (3)\tablenotemark{b} & $0.49 \pm 0.05$ \\  
NGC~4102 & $0.141 \pm 0.04$\tablenotemark{c} & \nodata\tablenotemark{a} & $0.08\pm0.08$  & $<0.12$ (3); $0.20\pm0.03$ (4)  & $0.08 \pm 0.08$ \\     
NGC~4214 & $0.31  \pm 0.03$\tablenotemark{d} & $0.31\pm0.03$\tablenotemark{d} & $0.55\pm0.05$  & $0.57 \pm 0.07$ (1) & $0.56 \pm 0.04$ \\  
NGC~4861 & $0.3\pm0.02$\tablenotemark{d} & $0.3\pm0.02$\tablenotemark{d,e} & \nodata & $0.36 \pm 0.03$ (1) & $0.36 \pm 0.03$ \\  
NGC~6217 & \nodata  & \nodata      & $0.39\pm0.04$  & \nodata                          & $0.39 \pm 0.04$ \\
 \enddata


\tablenotetext{a}{Without stellar continuum subtraction, the \HeIh\ line
cannot be measured.}
\tablenotetext{b}{Value ignored in computation of weighted mean.  Our
spectrum has much higher signal--to--noise.}
\tablenotetext{c}{For the $\HeIh$ feature, we directly summed
flux rather than fit a Gaussian.}
\tablenotetext{d}{\Brten\ was contaminated by an OH sky line 
in this galaxy, so the \HeIh/\Brten\ ratio we quote is scaled from 
\HeIh/\Breleven, assuming $[\Breleven / \Brten]_{case B} = 0.75$.}
\tablenotetext{e}{Only one measurement was made, so the quoted error
is an estimate based on the quality of the spectrum.}

\flushleft{
References-- (1) \citet{vr}; (2) \citet{dpj}; (3) \citet{doherty95}; (4) \citet{chad_thesis}.
}
\end{deluxetable}

\begin{deluxetable}{lccccc}
\rotate
\tablecolumns{6}
\tablewidth{0pc}
\tablenum{5}
\tablecaption{Model Predictions for Line Ratios.\label{tab-s99}}
\tablehead{
\colhead{line ratio} & \colhead{value for $t<5.5$~Myr,} & \colhead{peak value,} 
& \colhead{MS star at} & \colhead{WN at} & \colhead{WC at}\\
& \colhead{\Mup$=100$~\Msol} & \colhead{\Mup$=100$~\Msol} & \colhead{$\Teff=50,000$~K} & \colhead{$\Teff=100,000$~K} & \colhead{$\Teff=100,000$~K}\\
%
%
}
\startdata
$[\ArIII]/[\ArII]$  & $>0.6$   &  $13$     &   $14$     &   $25$   &   $35$\\
$[\NeIII]/[\NeII]$  & $>0.05$  &  $7$      &   $10$     &   $70$   &   $0.8$\\
$[\SIV]/[\SIII]$    & $>0.01$  &  $0.32$   &   $0.45$   &   $2.7$  &   $0.4$\\
$[\ArIII]/[\NeII]$  & $>0.2$   &  $2$      &   $2.5$    &   $12$   &   $0.65$\\
$[\SIV]/[\NeII]$    & $>0.02$  &  $2$      &   $4$      &   $90$   &   $0.9$\\
\HeIk/\Brgam  & $>0.5$   &  $0.8$    &   $0.75$   &   $0.25$ &   $0.4$\\
\HeIh/\Brten  & $>0.17$  &  $0.35$   &   $0.35$   &   $0.35$ &   $0.35$\\ 
\enddata

\flushleft{
Columns 2 and 3 list results from Starburst99/Cloudy models.
For comparison, column 4 gives peak line ratio values for a single--star 
model using a \Teff $= 50,000$~K main sequence star, and columns 5 and 6 list
peak values for single-star nebulae with Wolf--Rayet stars.
}
\end{deluxetable}


\begin{deluxetable}{llllll}
\tablenum{6}
\rotate
\tabletypesize{\small}
\tablecolumns{6}
\tablecaption{Testing optical \Teff\ diagnostics.  \label{tab-opt}}
\tablehead{
\colhead{Galaxy} & \colhead{\HeIh/\Brten}    &  \colhead{\HeIk/\Brgam}    &
\colhead{[\NeIII]/[\NeII]}  & \colhead{\Heopta/\Halpha}  &  \colhead{\Heoptc/\Hbeta}
           }
\startdata
%
NGC~3628 &  \nodata    & $<0.11$ (1)    & \nodata              &   not~det      &   not~det\\
M82~nucleus & $<0.16$ (2)    & \nodata     & $0.16\pm0.04$ (2) &   0.35         &   not det\\
NGC~6946 & \nodata     & $0.13\pm0.06$ (1) & $0.10$ (3)        &   not det      &   not det\\
NGC~972  & \nodata     & \nodata           &  $0.15$ (3)       &   $0.4$        &   not det\\
NGC~660  & \nodata     & $0.16\pm0.02$ (1) & \nodata           &   $0.5$        &   not det\\
NGC~4102 & $0.143\pm0.04$ & $0.08\pm0.08$  & \nodata           &   $0.51$       &   not det\\
NGC~6240 & \nodata     & $0.20\pm0.03$ (1) &  $<0.39$ (3)      &   not obs      &   not obs\\
NGC~3690 & \nodata     & \nodata           & $0.3$ to $0.7$\tablenotemark{a}~(3) & $0.28$  & not det\\
NGC~278  & \nodata     & \nodata           & $0.68$ (3)        &   not det      &   not det\\
\hline
NGC~3504 & $0.23\pm0.06$  & \nodata        &  \nodata             &   $0.36$       &   not det\\
NGC~4214 & $0.31\pm0.03$  & \nodata        &  \nodata             &   $0.83$       &   $0.36$\\
NGC~3077 & $0.46\pm0.084$ & \nodata        &  \nodata             &   $0.58$       &   $0.3$\\
\enddata

\tablenotetext{a}{Two pointings.}

\flushleft{
References:
(1) \citet{chad_thesis};
(2) \citet{fs-m82};
(3) \citet{thornley}.
All \Heopta/\Halpha\ and \Heoptc/\Hbeta\ flux ratios are from \citet{ho3}, 
dereddened by us, given as fractions of the saturated values ($0.014$ and 
$0.05$, respectively).  
Uncited near--infrared line ratios:  this paper.
``Not obs'' means galaxy was not observed by \citet{ho3}.
``Not det'' means galaxy was observed, but line was not detected.
The \HeIk/\Brgam\ ratio is only listed if $\le 0.2$, and therefore useful.
}
\end{deluxetable}

\begin{deluxetable}{lllll}
\tablenum{7}
\rotate
\tabletypesize{\scriptsize}
\tablecolumns{5}
\tablecaption{Comparison of Mid--IR and Near--IR \Teff\ diagnostics. \label{tab-midir}}
\tablehead{
\colhead{Object} & \colhead {$\HeIh/\Brten$} &  \colhead{$\HeIk/\Brgam$} &
\colhead{\emph{ISO} Mid--IR} & \colhead{Ground Mid--IR}\\}
\startdata
\objectname[M82]{M82--nucleus}   & $<0.13$  (1)      &  $0.48\pm0.07$ (1) & \nodata                            & [\ArIII]/[\NeII]$=0.054\pm0.02$ (2a)\\
                                 &                   &                    &                                    & [\SIV]/[\NeII]$=0.014\pm0.007$ (2a)\\ 
M82--E1                          & \nodata           &  \nodata           & \nodata                            & [\ArIII]/[\NeII]$=0.06\pm0.01$ (2)\\
                                 &                   &                    &                                    & [\SIV]/[\NeII]$=0.026\pm0.008$ (2) \\
M82--W1/B2                       & $0.22\pm0.06$ (1) & $0.52\pm0.02$ (1)  & \nodata                            & [\ArIII]/[\NeII]$=0.039\pm0.008$ (2)\\
                                 &                   &                    &                                    & [\SIV]/[\NeII]$=0.015\pm0.005$ (2)\\
M82--W2/B1                       & $0.22\pm0.05$ (1) & $0.55\pm0.02$ (1)  & \nodata                            & [\ArIII]/[\NeII]$=0.07\pm0.01$ (2)\\
                                 &                   &                    &                                    & [\SIV]/[\NeII]$=0.013\pm0.004$ (2)\\ 
M82--SWS $14\arcsec \times 20\arcsec$ &  $0.2\pm0.09$ (1a) & $0.51\pm0.03$ (1a) & [\NeIII]/[\NeII]$=0.16\pm0.04$ (1) & [\ArIII]/[\NeII]$=0.025\pm0.004$ (2b)\\
                                 &                   &                    & [\ArIII]/[\ArII]$=0.26\pm0.08$ (1)   & [\SIV]/[\NeII]$=0.007\pm0.0015$ (2b)\\
                                 &                   &                    & [\SIV]/[\SIII]$=0.11\pm0.04$   (1)   & \\
\objectname[NGC 253]{NGC~253}   &  $<0.15$ (3)	  &  $0.36\pm0.05$ (4) & [\NeIII]/[\NeII]$=0.06$ (5)        & [\ArIII]/[\NeII]$<0.3$ (6) \\
	                        &                 &                    &                                    & [\SIV]/[\NeII]$<0.3$ (6) \\

\objectname[NGC 4102]{NGC~4102} &  $0.14\pm0.04$  &  $0.08\pm0.08$     & \nodata			    & [\ArIII]/[\NeII]$<0.5$ (6) \\
                                &		  &		       &		                    & [\SIV]/[\NeII]$<0.2$ (6) \\
\objectname[NGC 6946]{NGC~6946} &  \nodata	  &  $0.13\pm0.06$ (4) & [\NeIII]/[\NeII]$=0.10$ (5)        & \nodata \\
\objectname[NGC 6240]{NGC~6240} & \nodata         & $<0.20\pm0.03$ (4) & [\NeIII]/[\NeII]$<0.39$ (5)        & \nodata \\
\objectname[He 2-010]{He~2--010} & $0.273\pm0.015$  & $0.53\pm0.03$ (11,8) &  \nodata	                    & [\SIV]/[\NeII]$=0.03\pm0.02$ (12) \\
                                &                   &                      &                                & [\ArIII]/[\NeII]$=0.08\pm0.03$ (12) \\
\tableline
\objectname[II Zw 40]{II~Zw40}	& $0.43\pm0.05$ (7) & $0.31\pm0.02$ (7) & [\NeIII]/[\NeII]$=12.0$ (5)       & [\SIV]/[\NeII]$>7$ (6) \\
                                &                 &                    &                                    & [\ArIII]/[\NeII]$>3$ (6) \\
\objectname[NGC 5253]{NGC~5253} & $0.19\pm0.03$ (8) & $0.48\pm0.01$ (9) & [\NeIII]/[\NeII]$=3.5$ (5)        & [\SIV]/[\NeII]$=4$ (10) \\
                                &                   &                   &                                   & [\ArIII]/[\NeII]$=0.7$ (10) \\
%
%
\enddata

\flushleft{
References:
(1) \citet{fs-m82}; 
(2) \citet{al}; 
(3) \citet{chad}; 
(4) \citet{chad_thesis}; 
(5) \citet{thornley}; 
(6) estimated from the spectra of \citet{roche}; 
(7) \citet{vanzi};
(8) \citet{vr}; 
(9) \citet{lpd};
(10) \citet{crowther}; 
(11) \citep{dpj};
(12) \citet{bkl}; uncited values are from this paper.
Positions in M82 are as follows: 
nucleus is defined at 9\fh51\fm43.6\fs, 69\fdg55\farcm00\farcs (1950) by (1) and (2); 
E1 is defined by (2) as the [\NeII] knot at 9\fh51\fm44\fs, 69\fdg55\farcm02\farcs\ (1950); 
W1 is defined by (2) as the [\NeII] knot at  9\fh51\fm42.5\fs,  +69\fdg54\farcm59\farcs\ (1950), 
	which corresponds to the \Brgam\ knot named B2 by (1);  
W2 is the [\NeII] knot found at  9\fh51\fm41.5\fs, 69\fdg54\farcm57\farcs\ (1950),
        which corresponds to the \Brgam\ knot named B1 by (1);
and SWS is the 14\farcs $\times$ 20\farcs\ aperture of SWS/ISO, centered on 9\fh51\fm42.2\fs, 
        69\fdg55\farcm00.7\farcs.  
Values labelled (1a) are the average over the 16\farcs $\times$ 10\farcs\ 3D field of (1), 
	which is similar in coverage to the SWS field.  
Values labeled (2a) are estimated from figure~7 of (2).
Values labelled (2b) are the average of knots W1 and W2 in (2), which provides similar 
	coverage to the SWS field.  
}
\end{deluxetable}


\clearpage


\clearpage


\clearpage


\clearpage


\clearpage


\clearpage


\clearpage


\clearpage

\clearpage

\clearpage

\clearpage

\clearpage



\end{document}